\documentclass[11pt]{article}
\usepackage{indentfirst}
\usepackage{graphicx}
\usepackage{amsmath,amsthm,amssymb,verbatim,url}
\usepackage{xifthen}
\usepackage{wasysym}
\usepackage{hyperref}
\usepackage{xifthen} % provides \isempty test
\hypersetup{colorlinks=true,urlcolor=blue}

\newcommand{\bv}{\begin{verse}}
\newcommand{\ev}{\end{verse}}
\newcommand{\be}{\begin{equation}}
\newcommand{\ee}{\end{equation}}
\newcommand{\bea}{\begin{eqnarray}}
\newcommand{\eea}{\end{eqnarray}}
\newcommand{\bq}{\begin{quotation}}
\newcommand{\eq}{\end{quotation}}

\newcommand{\myurl}[2][]{\ifthenelse{\isempty{#1}}{\url{#2}}{\href{#1}{\tt #2}}}

\begin{document}

\begin{center}
\LARGE {\bf Notwithstanding Bohr, the Reasons for QBism}\bigskip\bigskip\\
\Large Christopher A. Fuchs \bigskip \\
\normalsize Department of Physics, University of Massachusetts Boston \\ 100 Morrissey Boulevard, Boston MA 02125, USA \smallskip \\
and \smallskip \\
Max Planck Institute for Quantum Optics \\ Hans-Kopfermann-Strasse 1, 85748 Garching, Germany
\bigskip\\
%\large 12 January 2016
\end{center}

\bigskip\bigskip

\bq
\noindent \small
{\bf Abstract:}
Without Niels Bohr, QBism would be nothing.  But QBism is not Bohr.  This paper attempts to show that, despite a popular misconception, QBism is no minor tweak to Bohr's interpretation of quantum mechanics.  It is something quite distinct.  Along the way, we lay out three tenets of QBism in some detail:  1) The Born Rule---the foundation of what quantum theory means for QBism---is a normative statement.  It is about the decision-making behavior {\it any\/} individual agent should strive for; it is not a descriptive ``law of nature'' in the usual sense.  2) All probabilities, including all quantum probabilities, are so subjective they never tell nature what to do.  This includes probability-1 assignments.  Quantum states thus have no ``ontic hold'' on the world.  3) Quantum measurement outcomes {\it just are\/} personal experiences for the agent gambling upon them.  Particularly, quantum measurement outcomes are {\it not}, to paraphrase Bohr, instances of ``irreversible amplification in devices whose design is communicable in common language suitably refined by the terminology of classical physics.''  Finally, an explicit comparison is given between QBism and Bohr with regard to three subjects:  a) The issue of the ``detached observer'' as it arose in a debate between Pauli and Bohr, b) Bohr's reply to Einstein, Podolsky, and Rosen, and c) Bohr's mature notion of ``quantum phenomena.''  At the end, we discuss how Bohr's notion of phenomena may have something to offer the philosophy of William James: A physics from which to further develop his vision of the world---call it an {\it ontology\/} if you will---in which ``new being comes in local spots and patches.''
\eq

\bigskip

\section{Introduction: Bohr in the History of QBism}

\medskip

\begin{flushright}
\baselineskip=13pt
\parbox{4.1in}{\baselineskip=13pt\footnotesize
Notwithstanding the encouragement given to the pursuit of such inquiries by the great example of relativity theory which, just through the disclosure of unsuspected presuppositions for the unambiguous use of all physical concepts, opened new possibilities for the comprehension of apparently irreconcilable phenomena, we must realize that {\bf the situation met with in modern atomic theory is entirely unprecedented in the history of physical science}.}\\
\footnotesize --- Niels Bohr \cite[p.\ 19]{Bohr37-1}
\end{flushright}

It is good to know who your heroes in physics are, to have them near the top of your thoughts.  It can make the difference in a career.  There's a nice story about how John Wheeler would conduct faculty interviews at the University of Texas~\cite{Kimble04}.  Whereas other professors would ask the candidates about their research plans, teaching styles, funding successes, Wheeler would simply ask, ``Who are your heroes in physics and why?''

John Archibald Wheeler's hero in physics was Niels Bohr.  My own hero in physics was John Archibald Wheeler.  Wheeler liked to tell the story of applying for a National Research Council fellowship to work with Bohr~\cite{Bicak11}:  ``I can remember \ldots\ the words I put down in my application \ldots.  That was in 1934, very early 1934.  Why did I want to go to work with Bohr in Copenhagen?---It was because `he has the power to see further ahead in physics than any other man alive'.''  By the time I was on the physics scene in 1984 (50 years later!), I jumped at the opportunity to take a class with Wheeler because already then I sensed that ``he had the power to see further ahead in physics than any other man alive.''

I have never wavered on that, and I have never lost sight that Wheeler got to where he was, with his ``law without law''~\cite{Wheeler83c}, his ``physics as meaning circuit''~\cite{Wheeler86c}, his ``it from bit''~\cite{Wheeler88a}, through Bohr's great influence.  It's good to know who your heroes in physics are, but it is not because their words should be written in stone as timeless truths.  It is because they help set your agenda, give you orientation, and constellate your thoughts.  Rather than having the last word on anything, they serve the purpose of proposing the first words for the next wave of thinking.  So it was of Bohr with Wheeler, and then of Wheeler with me.

QBism~\cite{Fuchs10a,Fuchs14,Fuchs16,Fuchs02a} is a view on quantum mechanics born of this lineage, but taking its influences from quite a number of other directions as well:  the insistence of Asher Peres that quantum states are more properly compared to Liouville distributions than classical phase-space points; the personalist Bayesian view of probability of Bruno de Finetti and Frank Ramsey; Wolfgang Pauli's emphasis (in opposition to Bohr's de-emphasis) on the ``nondetachedness'' of the quantum observer; the great reset to the physics community's thinking that came with quantum information theory and quantum computing---i.e., that the quantum is not an expression of a {\it limitation} (as, say, the Heisenberg relation is usually thought of), but as an {\it enabler\/} of fantastic new informational capabilities unforseen in classical theory; and indeed the new technical formalisms arising with the development of quantum information. Furthermore, perhaps most broadly and deeply influential was the philosophy of William James and John Dewey, early American pragmatism.  But this list of outside influences only scratches the surface.

At first sight, its jumble of elements might appear more an eclectic mix than a purposeful recipe.  However, the list's accumulating ingredients for what ultimately became QBism all originated in one form or other through Wheeler, and Wheeler through Bohr.  They were each explored because there was something in the existing doctrine that called out for clarification or, more often than not, begged to be made sound and consistent.  Consistency in turn sometimes required new mathematical tools for its demonstration, and the reward on occasion was a theorem revealing some surprising aspect of quantum theory that helped make the founding conception that much more convincing.  And so it went, layer after layer.

A favorite example from early QBist thinking is something we called the quantum de Finetti representation theorem~\cite{Caves02b}.  Bohr didn't much talk about quantum states, but Wheeler did.  He thought one of the lessons of Bohr was that quantum states {\it just are\/} information.  This conviction, through Wheeler's influence, was world-changing, as it was one of the seeds that blossomed into the very field of quantum information theory. Wheeler would implore anyone within earshot to think about ``the information-theoretic color''~\cite{Wheeler89a} of quantum theory, and what a cast of students, postdocs, and visitors his message fell upon!\footnote{Student: Benjamin Schumacher (notions of qubit and quantum information, quantum coding).  Postdocs: David Deutsch (quantum Turing machine, first quantum algorithms), William Wootters (no-cloning theorem, quantum teleportation, entanglement measures, mutually unbiased bases), Wojciech Zurek (no-cloning, decoherence theory).  Visitors: Charles Bennett (theory of reversible computation, Maxwell's demon, quantum cryptography, quantum teleportation, entanglement theory), Richard Feynman (universal quantum computation, suggestion of quantum speed-up), Asher Peres (quantum teleportation, entanglement theory, contextuality proofs).} But if a quantum state is information, {\it whose information}?  Worse still, once the field of quantum information took shape, there was constant talk of the ``unknown quantum state.''  Unknown quantum states could not be cloned, but they could be teleported, protected with quantum error correcting codes, and even {\it revealed\/} through the technique of quantum-state tomography!  Imagine the frustration this phrase could cause for someone foundational-minded like myself.  If a quantum state just was information, how could it be unknown?  Surely it had to be known by someone or something?  More frustratingly still, how could it be revealed by making measurements on quantum systems, instead of only becoming available upon the asking of the knower?  If the concept of an ``unknown quantum state'' was genuinely necessary for the protocols of this new field, then quantum states looked to be intrinsic properties of quantum systems after all and Wheeler's starry-eyed information dreams a mistaken fantasy.

The answer to this conundrum had to be sought out in a technical statement, and this was what the quantum de Finetti theorem was about.  For, Bruno de Finetti had encountered a similar conundrum when he first espoused his subjectivist theory of probability in the early 1930s:  Statisticians for years had been speaking of how statistical sampling can reveal the ``unknown probability distribution.''  But from de Finetti's point of view, this makes as little sense as the unknown quantum state made for us.  What de Finetti's representation theorem established was that all this talk of an unknown probability was just that, talk.  Instead, one could show that there was a way of thinking of the resultant of statistical sampling purely in terms of a transition from prior subjective probabilities (for the sampler himself) to posterior subjective probabilities (for the sampler himself).  That is, every bit of statistical sampling from beginning to end wasn't about revealing a true state of affairs (the ``unknown probability''), but about the statistician's own states of information about a set of ``exchangeable'' trials, full stop.\footnote{For a pedagogical introduction, see Chapter 5 of \cite{stacey-thesis}.}  The quantum de Finetti theorem does the same sort of thing, but for quantum states.

Young and brash, I remember writing these words for an early draft of our quantum de Finetti paper:
\bq\small
What is a quantum state?  Since the earliest days of quantum theory, the predominant answer has been that it is a representation of the observer's knowledge of a system.  The quantum state
in itself has no objective reality.  This is a point of view the authors hold quite firmly.  Holding this view, however, does not require a concomitant belief that there is nothing left to learn in quantum foundations.  Indeed, it is quite the opposite.  Challenges arise every day, and the mental exercise required to arrest the conundrums builds an understanding and
problem-solving agility that {\it reading and rereading the founding fathers\/} cannot engender {\bf [citation]}.  One inevitably walks away from the challenge with a deeper sense for the physical content of quantum theory and a renewed confidence for tackling the most pressing questions of our day.  Questions as fundamental and distinct as ``Will a nonlinear extension of quantum mechanics be needed to quantize gravity?''\ and ``Which physical resources actually make quantum computation efficient?''\ start to feel tractable (and even connected) from this heightened point of view.
\eq
Just after the word ``engender'' in this passage, I made a citation that read as follows:
\bq\small
\noindent {\bf [citation]} For a sampling of the lengths that can be gone to in this regard, see: J.~Faye and H.~J. Folse, editors, {\sl Niels Bohr and Contemporary Philosophy}, (Kluwer, Dordrecht, 1994); H.~J. Folse, {\sl The Philosophy of Niels Bohr:~The Framework of Complementarity}, (North-Holland, Amsterdam, 1985); J.~Honner, {\sl The Description of Nature:\ Niels Bohr and the Philosophy of Quantum Physics}, (Oxford U. Press, Oxford, 1987); D. Murdoch, {\sl Niels Bohr's Philosophy of Physics}, (Cambridge U. Press, Cambridge, 1987); J.~Faye, {\sl Niels Bohr:\ His Heritage and Legacy.\ An Anti-Realist View of Quantum Mechanics}, (Kluwer, Dordrecht, 1991); S.~Petruccioli, {\sl Atoms, Metaphors and Paradoxes:\ Niels Bohr and the Construction of a New Physics}, translated by I.~McGilvray, (Cambridge University Press, Cambridge, 1993).
\eq

You see, my frustrations with Bohr as a springboard {\it for where I wanted physics to go\/} had long been festering.  In March 2001, I wrote to Henry Folse~\cite{Fuchs10}:
\bq\small
\noindent [W]hile I believe Bohr and his gang certainly started to point us in the right direction, I think we have a long, long technical way to go before we can claim a particularly deep understanding of the quantum structure.  Here's how I put it in exasperation to David Mermin once:\footnote{Indeed, the exasperation drips from the title of that earlier (1998) email to Mermin, ``The Bohrazine Shuffle''~\cite{Fuchs10}---a snarky allusion to the notion of a thorazine shuffle, ``the slow, slow, shuffling forward, usually while wearing slippers and a hospital gown, of a mental patient who has been rendered nearly catatonic by the tranquilizer thorazine.''\ \cite{UrbanDictionary}}
\bq\small
\noindent What's your take on this passage?  Can you make much sense of it?  What does he mean by ``providing room for new physical laws''?  What ``basic principles of science'' is he talking about?  What five pages of derivation are lying behind all this business?  It nags me that Bohr often speaks as if it is clear that the structure of quantum theory is derivable from something deeper, when in fact all the while he is taking that structure as given. \ldots\ %When did he ever approach an explanation of ``why complex Hilbert spaces''?  Where did he ever lecture on why we are forced to tensor products for composite systems?
It's a damned shame really:  I very much like a lot of elements of what he said, but as far as I can tell all the hard work is still waiting to be done.
\eq
The issue in my mind is {\it not\/} to {\it start\/} with complex Hilbert space, unitary evolution, the tensor product rule for combining systems, the identification of measurements with Hermitian operators, etc., etc., and {\it showing\/} that Bohr's point of view is {\it consistent\/} with that.  Instead it is to start with Bohr's point of view (or some variant thereof) and see that the {\it precise\/} mathematical structure of quantum theory {\it must\/} follow from it.  Why complex instead of real?  Why unitary, rather than simply linear?  Indeed, why linear?  Why tensor product instead of [direct] sum?  And, so on. When we can answer {\it these\/} questions, then we will really understand complementarity.

I'm banking my career on the idea that the tools and issues of quantum information theory are the proper way to get at this program.
\eq
But the truth is, I myself was reading and rereading Bohr even as I wrote these things \ldots\ and reading and rereading Pauli \cite{Pauli94,Meier01,Laurikainen88} and Heisenberg \cite{Heisenberg71,Heisenberg74,Heisenberg62,Heisenberg52} and Rosenfeld \cite{Rosenfeld79} and von Weizs\"acker \cite{Weizsaecker80} \ldots\ and Faye \cite{Faye91} and Folse \cite{Folse85} and Plotnitsky \cite{Plotnitsky94,Plotnitsky06}.  I tried---I tried really, really hard---to find some angle, some way of understanding Bohr and the other Copenhageners that would convince me that they had given a consistent account of quantum theory (consistent to my standards at least).  I never got there.  Their doctrines just never fully made sense to me.

Take Bohr himself: To this day, quite a number of his statements remain near complete mysteries to me.  What means ``irreversible amplification''?  How would I ever know if I had something so amplified?  (Asher Peres used to say, ``Give me enough money, and I shall reverse the world; it's just a question of effort.'')  Or, take when Schr\"odinger implored Bohr in 1935~\cite{SchroedingerLetter35}, ``[C]ouldn't you make this point completely clear in the more detailed paper you announce \ldots: Why do I [Bohr] emphasize again and again that according to the very nature of a measurement, it can only be interpreted classically?''  Bohr's response?  Nothing more than his usual one-sentence statement on the matter~\cite{BohrLetter35}:
\bq\small
\noindent My emphasis of the point that the classical description of experiments is unavoidable amounts merely to {\it the seemingly obvious fact that the description of any measuring arrangement must, in an essential manner, involve the arrangement of the instruments in space and their functioning in time, if we shall be able to state anything at all about the phenomena.}\footnote{Compare this to Bohr's more mature formulation of 14 years later~\cite[p.\ 39]{Bohr49-1}, ``[I]t is decisive to recognize that, {\it however far the phenomena transcend the scope of classical physical explanation, the account of all evidence must be expressed in classical terms}.  The argument is simply that by the word ``experiment'' we refer to a situation where we can tell others what we have done and what we have learned and that, therefore, the account of the experimental arrangement and of the results of the observations must be expressed in unambiguous language with suitable application of the terminology of classical physics.'' In the field of quantum computing, people often tell each other of what they have done by drawing a quantum circuit, for instance for the Shor factoring algorithm.  For fun, do a Google Image search on ``quantum circuit'' to see what I am talking about.}
\eq
Well, it certainly wasn't ``seemingly obvious'' to me.  Consider a sensation of warmth I might feel from holding a sample of plutonium.  Is that describable in classical terms?  Must it be?  Isn't the sensation more basic than classical physics itself anyway?  These things bothered me.  Was Bohr reticent to say more simply because there was nothing there to say?  When there's genuinely something there, one can usually hit it from every angle imaginable.  Finally, maybe my most basic frustration was with Bohr's repeatedly invoked term ``the quantum postulate'' \cite[p.\ 53]{Bohr27-1}:
\bq\small
\noindent
Notwithstanding the difficulties which, hence, are involved in the formulation of the quantum theory, it seems, as we shall see, that its essence may be expressed by the so-called quantum postulate, which attributes to any atomic process an essential discontinuity, or rather individuality, completely foreign to the classical theories and symbolized by Planck's constant of action.
\eq
One would think this notion something so basic, so important for all quantum theoretical considerations, its meaning should jump out with its very statement.  ``The reason we have the formalism is because of the quantum postulate and the quantum postulate is this: [{\it fill in the blank in a way I can understand\/}].''  I searched and searched to find an account of the notion that was independent of already having the full-blown formalism.  I never found it. \label{QuantPost}

But then, why did I not simply give up on Bohr and cut my losses?  Why did I not embrace Bohmian mechanics, spontaneous collapse, or the Everettian multiverse for a more satisfactory account of quantum theory?  The followers of these interpretations certainly claim them to be consistent, and wasn't that what I was all about, consistency?  At the time I started down this long path, I had never heard of the philosopher William James, but with hindsight I would say he hit it on the mark in his 1907 Lowell Lectures~\cite{James1907},
\bq\small
The history of philosophy is to a great extent that of a certain clash of human temperaments.  Undignified as such a treatment may seem to some of my colleagues, I shall have to take account of this clash and explain a good many of the divergencies of philosophies by it.  Of whatever temperament a professional philosopher is, he tries, when philosophizing, to sink the fact of his temperament. Temperament is no conventionally recognized reason, so he urges impersonal reasons only for his conclusions.  Yet his temperament really gives him a stronger bias than any of his more strictly objective premises.  It loads the evidence for him one way or the other \ldots\ just as this fact or that principle would.  He {\it trusts\/} his temperament.  Wanting a universe that suits it, he believes in any representation of the universe that does suit it.
\eq
It was temperament.  James, in a different essay~\cite{James1879} elaborates,
\bq\small\noindent
Why does Clifford fearlessly proclaim his belief in the conscious-automaton theory, although the `proofs' before him are the same which make Mr.\ Lewes reject it?  Why does he believe in primordial units of `mind-stuff' on evidence which would seem quite worthless to Professor Bain?  Simply because, like every human being of the slightest mental originality, he is peculiarly sensitive to evidence that bears in some one direction. It is utterly hopeless to try to exorcise such sensitiveness by calling it the disturbing subjective factor, and branding it as the root of all evil. `Subjective' be it called!\ and `disturbing' to those whom it foils!  But if it helps those who, as Cicero says, ``vim naturae magis sentiunt'' [feel the force of nature more], it is good and not evil. Pretend what we may, the whole man within us is at work when we form our philosophical opinions. Intellect, will, taste, and passion co-operate just as they do in practical affairs \ldots.
\eq
The moral he drew was persuasive:
\bq\small
\noindent [I]n the forum [one] can make no claim, on the bare ground of his
temperament, to superior discernment or authority.  There arises thus
a certain insincerity in our philosophic discussions:  the potentest
of all our premises is never mentioned.  I am sure it would
contribute to clearness if in these lectures we should break this
rule and mention it, and I accordingly feel free to do so.
\eq
And as such, I accordingly feel free to do so myself!  Wheeler's vision of a world built upon ``law without law,'' a world where ``the big bang is here''\footnote{This phrase comes from a remarkable conversation recorded in \cite{Wheeler82c} between the Reverend Richard Elvee and John Wheeler.  Elvee asked, ``Dr.\ Wheeler, who was there to observe the universe when it started?  Were we there?  Or does it only start with our observation? Is the big bang here?''  Wheeler answered, ``A lovely way to put it---`Is the big bang here?' I can imagine that we will someday have to answer your question with a `yes.' \ldots  Each elementary quantum phenomenon is an elementary act of `fact creation.' That is incontestable. But is that the only mechanism needed to create all that is?  Is what took place at the big bang the consequence of billions upon billions of these elementary processes, these elementary `acts of observer-participancy,' these quantum phenomena?  Have we had the mechanism of creation before our eyes all this time without recognizing the truth? That is the larger question implicit in your comment. Of all the deep questions of our time, I do not know one that is deeper, more exciting, more clearly pregnant with a great advance in our understanding.''} and the observer is a {\it participator\/} in the process, attracted me like no other physical ideas I had ever heard before.  They were in fact the potentest of all {\it my\/} premises.  Whenever, in my younger days, I would read a passage of Wheeler's like this one~\cite{Wheeler82c},
\bq\small
How did the universe come into being? Is that some strange, far-off process beyond hope of analysis?  Or is the mechanism that comes into play one which all the time shows itself? Of all the signs that testify to ``quantum phenomenon'' as being the elementary act and building block of existence, none is more striking than its utter absence of internal structure and its untouchability.  For a process of creation that can and does operate anywhere, that is more basic than particles or fields or spacetime geometry themselves, a process that reveals and yet hides itself, what could one have dreamed up out of pure imagination more magic and more fitting than this?
\eq
I would get absolutely giddy with excitement.  And, if Wheeler {\it thought\/} these things were suggested by Bohr's understanding of ``the elementary quantum phenomenon,'' then to my subliminal mind, Bohr {\it had to be\/} salvaged somehow, someway.  It made all the excruciating effort worth it.

Of course, there was always a certain sense of danger in following Wheeler, and I suppose that was temperament too.  Stephen Brush aptly named the issue ``Wheeler's dilemma''~\cite{Brush80}:
\bq\small
\noindent [H]ow can one maintain a strong version of
the Copenhagen Interpretation, in which the observer is inextricably
entangled with that which is observed, while at the same time denying
that our consciousness affects that which we are conscious of---and
thus accepting the possibility of telekinesis and other psychic
effects?  For Wheeler himself there is no dilemma at all; one simply
has to recognize ``the clear distinction between (1) the strange but
well verified and repeatable features of quantum mechanics and (2) the
pseudo-scientific, non-repeatable and non-verified so-called extra
sensory perception.''  But Wheeler's own views are
likely to strike a non-physicist as being just as bizarre as those of
the parapsychologists he deplores.  Indeed, no one has yet formulated
a consistent worldview that incorporates the Copenhagen Interpretation
of Quantum Mechanics while excluding what most scientists would call
pseudo-sciences---astrology, parapsychology, creationism, and
thousands of other cults and doctrines.
\eq
I took Wheeler's dilemma quite seriously.  So had Einstein, long before Wheeler ever came upon the scene.  In the ``Reply to Criticisms'' in the Schilpp volume~\cite{Einstein49}, Einstein writes,
\bq\small
I close these expositions, which have grown rather lengthy, concerning the interpretation of quantum theory with the reproduction of a brief conversation which I had with an important theoretical physicist.\footnote{That important theoretical physicist was almost surely Wolfgang Pauli. An evidential stack can be found in \cite{Laurikainen88} and \cite{Meier01}.  Relatedly, Pauli himself writes of Einstein~\cite{Pauli58a}:  ``We often discussed these questions together, and I invariably profited very greatly even when I could not agree with Einstein's views. `Physics is after all the description of reality,' he said to me, continuing, with a sarcastic glance in my direction, `or should I perhaps say physics is the description of what one merely imagines?'  This question clearly shows Einstein's concern that the objective character of physics might be lost through a theory of the type of quantum mechanics, in that as a consequence of its wider conception of objectivity of an explanation of nature the difference between physical reality and dream or hallucination become blurred.''}  He:  ``I am inclined to believe in telepathy.''  I:  ``This has probably more to do with physics than with psychology.''  He: ``Yes.'' ---
\eq
Wheeler's dilemma had to be tamed somehow, someway, just as Bohr had to be salvaged.  This became part of the challenge.

Later I would discover quite by accident~\cite{Fuchs11a} the philosophies of William James, John Dewey, and F.~C.~S. Schiller, i.e., early American pragmatism; the tychism of Charles Sanders Peirce;\footnote{I draw a distinction between Peirce and the other three pragmatists on purpose.} the somewhat allied French philosophies of Charles Renouvier, \'Emile Boutroux, Henri Bergson, and Jules Lequyer; and the Italian pragmatisms of Giovanni Papini and Bruno de Finetti.  When I came upon these things, for the first time I realized that Wheeler's thinking, though grounded in modern physics, was just the tip of an iceberg.  There was a vast wealth of thought in these directions that, to my mind, had simply come before its time and remained thus (but only thus) sadly disconnected from physics.  Perhaps these philosophies held the key for unlocking Bohr, and quantum theory the key for unleashing their own potentials!

So far, I have presented this story from the side of my own motivations---specifying only what was driving me and nagging me idiosyncratically.  That description puts the emotions in place, but does little to fill the gap of how QBism actually developed.  Truth is, I would have never gotten anywhere by myself; the project of QBism was a much larger collective enterprise.  New thinking was needed far beyond better {\it Bohr interpretation}. A place and a time can make all the difference, and that place and time was Albuquerque, New Mexico in the mid-1990s. There, three Bayesians, Carlton Caves, R\"udiger Schack, and Howard Barnum, and one frequentist/propensitist about physical probabilities (namely, me) spent two years debating it out day-by-day as to how two conceptually distinct, yet formally identical, notions of probability could make their way into quantum theory.  All the while we were sipping from the cup of the newly brewing quantum information theory.

This is where the original B in QBism---namely Bayesianism---came into the picture.  It wasn't that two notions of probability were needed, but that one should doggedly stick with the Bayesian notion all the way through and see what came of it.  The key guidance was in the writings of E.~T. Jaynes.  It was the crucial input for breaking the impasse on Bohr~\cite{Caves02a} and breaking the impasse of Wheeler's dilemma.  It still guides the uncompleted project of QBism today.  Jaynes wrote~\cite{Jaynes90},
\bq\small
[O]ur present QM formalism is not purely epistemological; it is a peculiar mixture describing in part realities of Nature, in part incomplete human information about Nature---all scrambled up by Heisenberg and Bohr into an omelette that nobody has seen how to unscramble.  Yet we think that the unscrambling is a prerequisite for any further advance in basic physical theory.  For, if we cannot separate the subjective and objective aspects of the formalism, we cannot know what we are talking about; it is just that simple.
\eq
Furthermore~\cite{Jaynes89},
\bq\small
\noindent It seems that, to unscramble the epistemological probability statements from the ontological elements we need to find a different formalism, isomorphic in some sense but based on different variables; it was only through some weird mathematical accident that it was possible to find a variable $\psi$ which scrambles them up in the present way.
\eq
It hit our heads like a board.  ``Separate the subjective from the objective.  That's it!  That must be our goal!''  Wheeler might well have been right that the observer is a {\it participator}---``an actor and a coefficient of the truth'' as William James would put it---but if so, that would have to be a hard-won result of the formalism, not a hopeful declaration.  Until quantum theory could be cleaned up, there was still too much confusion to know exactly where or how a full-throated notion of participator might fit in.  So it was a question of step-by-step.  The starting point was, along with Jaynes, that {\it the objective\/} could never reside in the probabilities themselves; that was for the subjective.  The objective must be somewhere else in the theory.  And that defined our research program.

Here is the way I put it in one of its earliest formulations~\cite{Fuchs02a} (it's not exactly language I would use today, with too much talk of ``information'' in place of ``degrees of belief,'' but it conveys the zeitgeist):
\bq\small
This, I see as the line of attack we should pursue with relentless consistency:  The quantum system represents something real and independent of us; the quantum state represents a collection of subjective degrees of belief about {\it something\/} to do with that system (even if only in connection with our experimental kicks to it). The structure called
quantum mechanics is about the interplay of these two things---the subjective and the objective.  The task before us is to separate the wheat from the chaff.  If the quantum state represents subjective information, then how much of its mathematical support structure might be of that same character?  Some of it, maybe most of it, but surely not all of it.

Our foremost task should be to go to each and every axiom of quantum theory and give it an information theoretic justification if we can.  Only when we are finished picking off all the terms (or combinations of terms) that can be interpreted as subjective information will we be in a position to make real progress in quantum foundations.  The raw distillate left behind---minuscule though it may be with respect to the full-blown theory---will be our first glimpse of what quantum mechanics is trying to tell us about nature itself.
\eq

Once this idea was taken on board, then it was just a matter of consistency, consistency all the way down!  For instance, suppose one accepts a fully subjectivist or personalist Bayesian account of all quantum probabilities---including probabilities 0 and 1---then what must follow for the Hermitian operators we use for describing quantum measurements?  Must they too have a personalist, subjective character?  Or consider the paradox of Wigner's friend.  If one accepts a purely personalist account of both quantum states {\it and\/} quantum operators, what does that entail for the very nature of quantum measurement outcomes?  Could they really consistently be ``objectively available for anyone's inspection,'' as Pauli~\cite{Pauli58a} had said of them?  Or was one forced to dip back into the philosophies of William James and John Dewey, with quantum measurement outcomes taken more rightfully to be ``personal experiences'' rather than depersonalized fact?  Yet, if all these subjectivities were accepted, how could there be any place left for objectivity to be located at all?  Perhaps it was only in the {\it relation\/} expressed by the Born rule, not in the individual elements within the relation?

These questions express just a few of the considerations that had to be struggled through:  It took a village (Caves, Schack, Mermin, Peres, Appleby, von Baeyer, Barnum) situated in a hostile landscape (the large number of philosophers of science who helped sharpen QBism by fighting every bit of it tooth and nail) to get it done.  In the end, we were left with an interpretation of quantum theory that was very different from Bohr's, though the roots have always remained visible.  Better yet, the fruit is what Wheeler would have probably most wanted to preserve anyway: For, Bohr's ``elementary quantum phenomenon,'' in new guise, strikes out with more vitality than ever.

This essay is about uncovering some of the said roots for more careful inspection---both the roots that are alive from the perspective of QBism, and notwithstanding Bohr, the roots that QBism considers dead.  In content, this might be considered a supplement to David Mermin's excellent essay, ``Why QBism is Not the Copenhagen Interpretation and What John Bell Might Have Thought of It''~\cite{Mermin14f}.  The structure of what remains is this.  In Section \ref{Tenets}, I describe what QBism actually entails as a point of view on quantum theory.  Particularly I lay stress on three of its most important tenets.  Armed with these, in the section that follows, I show how QBism responds to three perennial issues in Bohr scholarship: how QBism sides more with Pauli than Bohr on the notion of ``detatched observer'' but still goes further, how QBism's response to EPR differs fundamentally from Bohr's, and how QBism fits with Bohr's notion of phenomena.  Finally in Section \ref{Future}, I come back to the vistas physics might encounter from the top of the tree first planted by Bohr and grafted by Pauli and Wheeler. Though he had no right to get so high without the safety rope of physics, it seems William James might already be there waiting to point at some sights in the distance.  In an Appendix, I give an exhaustive collection of the Bohr quotes that inspired this essay's title. Keeping on the lookout for these became the amusement that ``got me through the night'' when Bohr's writings were the most repetitive.

\section{What is QBism?}
\label{Tenets}

QBism, initially an abbreviation for Quantum Bayesianism, is now pretty much a stand-alone term for the point of view of quantum theory sketched historically in the Introduction.\footnote{Detailed accounts of QBism can be found in \cite{Fuchs10a,Fuchs14,Fuchs16,Fuchs02a,Mermin14f,Fuchs13a} and especially \cite{Varenna17}. I might also recommend Richard Healey's very accurate article in the Stanford Encyclopedia of Philosophy~\cite{Healey16} and sections III and IV of Hans Christian von Baeyer's popularization of the subject~\cite{vonBaeyer16}.}  David Mermin put the distinction very flatteringly in \cite{Mermin13}, ``I prefer [the term QBism] because [this] view of quantum mechanics differs from others as radically as cubism differs from renaissance painting, and because I find the term `Quantum Bayesianism' too broad.'' The key point of the second half of his sentence is that there are many varieties of ``objective Bayesianism,'' and even some mildly subjective ones~\cite{Good83}, to which QBism does not subscribe~\cite{Stacey16}.  For a while we toyed with saying that the B stood for the B in Bruno de Finetti, the founder of the variety of subjective or personalist Bayesianism that QBism {\it does\/} subscribe to, but that was certainly tongue-in-cheek.  If there were a proper word in the history of philosophy that the B genuinely stands for, one would have to say it is the rather unattractive term ``bettabilitarianism'' invented by Oliver Wendell Holmes Jr.\ to describe his view of the universe.  Holmes wrote~\cite[pp.~251--253, his {\it italics}, my \underline{underlining}]{Holmes29}:
\bq\small
\noindent Chauncey Wright a nearly forgotten philosopher of real merit, taught me when young that I must not say {\it necessary\/} about the universe, that we don't know whether anything is necessary or not.  So I describe myself as a {\it bet\/}tabilitarian.  I believe that we can {\it bet\/} on the behavior of the universe in its \underline{contact} with us.
\eq
From this point of view, a gambling agent who adopts quantum theory becomes a ``better bettabilitarian''~\cite{Varenna17}.  He is better able to thrive in the world in which he is immersed---a world whose characteristics cause the quantum formalism to become an advisable addition to his probability calculus.

Three characteristics set QBism apart from other existing interpretations of quantum mechanics. The first is its crucial reliance on the mathematical tools of quantum information theory to reshape the look and feel of quantum theory's formal structure. The second is its stance that two levels of radical ``personalism'' are required to break the interpretational conundrums plaguing the theory. The third is its recognition that with the solution of the theory's conundrums, quantum theory does not reach an end, but is the start of an adventure: QBism is a continuing project.

\begin{figure}[h]
\begin{center}
\includegraphics[height=4.0in]{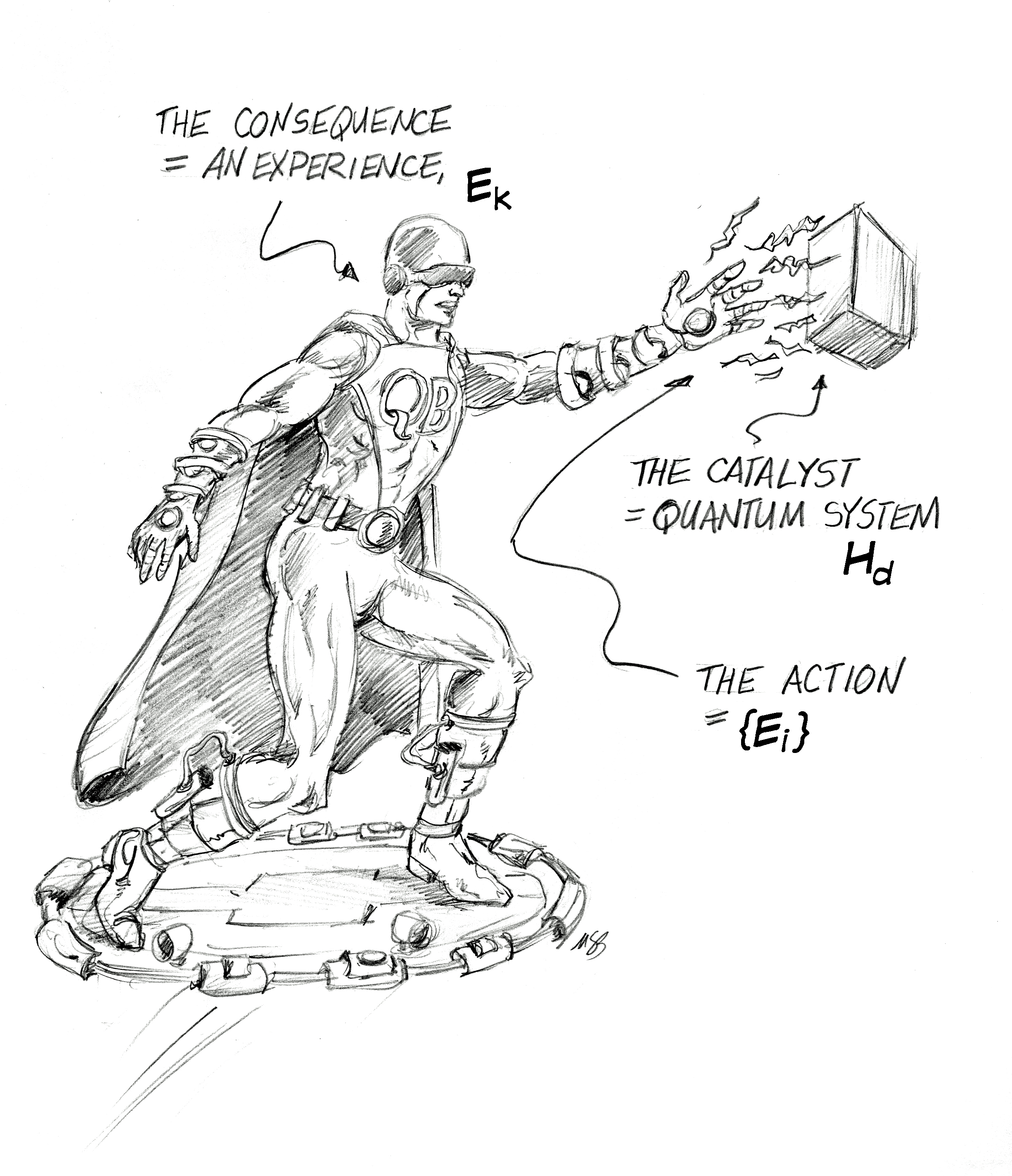}
\caption{``May not the {\it creatia\/} of a quantum observer's actions likewise be such additions to the universe as to enhance its total value?  And on this view, is not the QBist quantum observer---the agent---a kind of superhero for the universe as a whole, making {\it extra\/} things happen wherever and whenever he is called to duty?''~\cite{Varenna17} {\footnotesize (Drawing of Captain QBism courtesy of \myurl[http://www.drgreatart.com/]{Mark Staff Brandl}.)}}
\end{center}
\end{figure}

The two levels of personalism refer to how the ``probabilities'' and ``measurement events'' of quantum theory are to be interpreted. With regard to quantum probabilities, QBism asserts that they are to be interpreted as genuinely personal, Bayesian degrees of belief. This is the idea that probability is not something out in the world that can be right or wrong, but a personal accounting of what one expects. The implications of this are deep, for one can see with the help of quantum information theory that it means that quantum states, too, are not things out in the world. Quantum states rather represent personal accounting, and two agents speaking of the same quantum system may have distinct state assignments for it. In fact, there are potentially as many quantum states for a system as there are agents interested in considering it.\footnote{Just to give a very clear-cut example of how QBism has moved beyond its historical roots, it is worth recording a fascinating passage Amanda Gefter unearthed in one of John Wheeler's notebooks~\cite{Gefter16}.  Dated 14 March 1976, Wheeler wrote, ``In seminar Wed.\ 10 Mar (7:30 -- 9pm) Andrew Redfield claimed the wave function represents the state of our information about the electron (I agree) but that there's a different wave function for each observer because each observer has different information about that electron (I disagree).  If quantum mechanics is a magic clue to the great mysteries, it ought to be especially so on this aspect of observership.''}

The second level of personalism appears with the meaning of a quantum-measurement outcome. On this, QBism holds with Pauli (and against Bohr) that a measurement apparatus must be understood as an extension of the agent himself, not something foreign and separate. A quantum measurement device is like a prosthetic hand, and the outcome of a measurement is an unpredictable, undetermined ``experience'' shared between the agent and the external system. The latter part of this we see as the ultimate message of the Bell and Kochen-Specker theorems and their variants~\cite{Cabello16,Ben-Menahem97}---that when an agent reaches out and touches the world, a little moment of creation occurs in response. Quantum theory, thus, is no mirror image of what the world is, for ``there is no one way the world is;'' it is ``still in creation, still being hammered out''~\cite[p.~742]{FuchsOx}.  Rather the theory should be seen as a ``user's manual'' that any agent can adopt for better coping with the world external to him. The agent uses the manual to help guide his little part and participation in the world's ongoing creation.

Figure 1 attempts to capture the key concerns of QBism in a single diagram.  In contemplating a quantum measurement, one makes a conceptual split in the world: One part is treated as an agent, and the other as a kind of reagent or catalyst (in any case, something that brings about a change in the agent). In older terms, the former is an observer and the latter a quantum system modeled by a Hilbert space ${\cal H}_d$ of some finite dimension $d$.\footnote{\label{FootNoteX} In the development here, we will only be concerned with finite-dimensional quantum-state spaces, but there is nothing essential about this, except that the formalism for Section~\ref{FirstTenet} becomes much cleaner if we stick with finite dimensions.  Everything can be done again for infinite-dimensional Hilbert spaces but at the cost of equations which aren't nearly so pretty.  At times we have wondered whether this is a hint that quantum mechanics should more properly be restricted to finite spaces \cite{Fuchs10a,Varenna17}, but this is not the place to discuss it.}  A quantum measurement consists first in the agent taking an {\it action\/} on the quantum system. The action is formally captured by a set of measurement operators $\{\hat E_i\}$ (a positive-operator valued measure; see Section \ref{FirstTenet}). The action leads generally to an incompletely predictable consequence, a particular personal {\it experience\/} $E_k$ drawn from the set. The action is identified with the set, the experience with a single element within the set.  Furthermore, as anything in the agent's external world may be considered a quantum system, an action can be anything from running across the street at L'Etoile in Paris (and gambling upon one's life) to a sophisticated quantum information experiment (and gambling on the violation of a Bell inequality).

The quantum state $|\psi\rangle$ for the system makes no appearance because it is inside the agent's head: It only captures his degrees of belief concerning the consequences of his actions upon the system, and in contrast to the system itself, has no existence in the external world. If the agent were to go poof, the quantum state would go poof too.  Quantum states are not part of the scaffolding of the world; only quantum systems are.

Notice also the eye shades on the agent:  This is meant to subtly convey that quantum measurement is not a passive observation of what is out there, in the motif of an eye looking upon the world, but an action upon the world that aids it in releasing its potential. The measuring devices are depicted as mechanically enhanced hands to make it clear that they should be considered an integral part of the agent himself. The sparks between the measurement-device hand and the quantum system represent the idea that the consequence of each quantum measurement is a unique creation within the previously existing universe.  Wolfgang Pauli characterized this picture as a ``wider form of the reality concept''~\cite{Fuchs16} than that of Einstein's, which he labeled ``ideal of the detached observer.''

Let us now put more detail into this diagram by considering three fundamental tenets of QBism.

\subsection{First Tenet: Quantum Theory Is Normative, Not Descriptive}
\label{FirstTenet}

If for QBism quantum theory is not a direct {\it description\/} of reality as it is, then what does it mean for a physicist or an experimenter to ``accept quantum theory''?

By way of warming up for a proper answer, imagine first an encounter one could easily have in any city in the U.S.  A man on the street asks, ``Are you a Christian?''  I say, ``What does that mean?''  He tries again, ``Do you accept Jesus Christ as your personal Lord and Savior?''   Ah, that's where he is going. What the latter is really about is a kind of code for, ``Are you {\it trying\/} to live a life of the type that Jesus taught?  Do you {\it try\/} your best to turn the other cheek?  Do you {\it try\/} your best to forgive others of their sins?  Do you {\it try\/} your best to treat others as you would have them treat you?''  There is a whole corpus of ideas in this compact code, and they are all about {\it trying}.  In this sense Christianity is a {\it normative\/} doctrine:  It is something one aims for, but perhaps never achieves.  QBism says something similar for quantum theory.

``Do you accept quantum theory?''  For a QBist, this translates to:  Do you accept the Born Rule as a personal normative aim in aid of your decisions?  ``What does this question even mean?''\ a reader might ask. ``To use the Born Rule means to take the quantum state implied by the preparation procedure and the projections associated with the measurement device and put them together to calculate probabilities for the measurement outcomes.''  The state and the projections, {\it on this view}, are logically prior to the probabilities calculated from them. They are entities of a distinct character from probability; so no wonder one might have gotten into the habit of thinking of them as something more fundamental than the probabilities they give rise to.  Facts of nature even.   ``You're not going to tell me that the ground state of intergalactic hydrogen is anyone's degree of belief,''\ Leslie Ballentine once said at a conference~\cite{Ballentine03}.  {\it Well, in QBism it is exactly that!}  It may be a widely shared ``degree of belief'' by all those who know enough quantum theory to formulate it, but conceptually a belief it remains.  What is important {\it and objective\/} in QBism is {\it the method\/} that gets us to such a belief.

An example of what we have in mind could go something like this~\cite{Caves07}.  Take a spin-1/2 particle.  Suppose an agent contemplates acting on it in such a way that a $\hat\sigma_x$ measurement is made.  He asks himself how he would gamble on those potential outcomes---i.e., what his expectation value $\langle\sigma_x\rangle$ might be.  He next contemplates the same of a $\hat\sigma_y$ measurement, and then of a $\hat\sigma_z$ measurement.  This {\it mental process\/} supplies the agent with three distinct probability distributions that {\it a priori\/} need bear no relationship to each other.

But then suppose the agent contemplates how he would gamble upon the outcomes of a spin measurement $\hat\sigma_n$ in some fourth direction?  It is here that the Born Rule gives some guidance---for the Born Rule is an addition to the raw probability calculus. Using it signifies that the context of the probability assignments is not some general setting, but that of various hypothetical (and complementary) quantum measurements the agent could perform on the given system.  The three first probabilities, being {\it informationally complete}, specify a unique density operator $\hat\rho$ which the agent is free to {\it imagine\/} as mathematically generating the probabilities in the first place. In other words, the three equations $\langle\sigma_x\rangle={\rm tr}\,\hat\rho\hat\sigma_x\,$, $\langle\sigma_y\rangle={\rm tr}\,\hat\rho\hat\sigma_y\,$, $\langle\sigma_z\rangle={\rm tr}\,\hat\rho\hat\sigma_z\,$, where tr denotes the trace,  will have a unique solution for the variable $\hat\rho$.  This density operator then, once again through the Born Rule, will lead uniquely to the fourth set of probabilities, and consequently a fourth expectation value $\langle\sigma_n\rangle={\rm tr}\,\hat\rho\hat\sigma_n\,$.  That is, once the three expectations are set, the fourth has no freedom at all.  In this sense, the expectations come first and last; the density operator is only a construct in the middle.

However, this is all a mental game for the agent.  Probabilities are not given to him on a plate.  Ultimately, they are guesses that he makes after trying to take as many considerations into account as he has time or mental capacity for.\footnote{This is perhaps a good place to emphasize a key distinction between the personalist notion of probability so emphasized by QBism and the objective Bayesian notion espoused by E. T. Jaynes~\cite{Jaynes03}.  A recent commentator put the issue like this~\cite{Mueller16}: ``I don't buy the difference between `objective Bayesianism' and (de Finetti-type) `personalist Bayesianism' in general; the distinction seems artificial to me.  According to the latter viewpoint, different agents can assign different probabilities because they have different backgrounds and experiences.  Fine.  But the thing is: For {\it fixed given\/} background and experiences (including one's knowledge of one's name, one's memories, previous probability assignments, really everything), the assignment should be objective.  That is, there should be a normative assignment of probabilities
$P(y|x)$
where $y$ encodes an agent's future experiences, and $x$ encodes the totality of {\it all\/} the agent's past experiences, everything she knows and sees and remembers right now. This is in line with {\it both\/} views on Bayesianism, as far as I can see. It just follows from the fact that `you only have what you have' (which is $x$), and you {\it must\/} assign some probability.''  But in fact, this is exactly where QBism differs on things:  An agent {\it never\/} has $x$ as defined here, or at least an $x$ he can reach.  Even with respect to the single agent, the probability assignment of choice is not unique.  In the end, a probability assignment always involves a ``leap of faith'' in the sense of William James's ``will to believe'' doctrine~\cite{James1879,James1896}, which van Fraassen underlines well in his essay ``Belief and the Will''~\cite{Fraassen84}.  For the mesh of beliefs simply can never be made complete and exhaustive.  If the world itself is not ``complete and exhaustive''---as the very ontology QBism toys with indicates---why should anyone's mesh of beliefs be?  QBism thus interprets {\it probability theory as a whole\/} in the normative sense, but not so for any of the provisionary probability assignments within it.}
If after calculating the fourth set of probabilities, he finds them so absurd, so unreflective of how he thinks he should gamble on the outcomes of the fourth measurement, he might go back and completely readjust his expectations for $\hat\sigma_x$ or $\hat\sigma_y$ or $\hat\sigma_z$ or any combination of the three.  For instance, he might do this based on what he {\it more genuinely believes\/} of a measurement of $\hat\sigma_n$ (perhaps he has more experience making measurements in that odd direction because his lab table is a little crooked or something).  Nothing is sacred except that he should {\it strive\/} to satisfy the Born Rule for all four probabilities.  Which is to say, the agent should {\it strive\/} to find a single quantum state $\hat\rho$ assignment that encapsulates all the relevant probabilities.  As it is all a mental game, however, as the probabilities are adjusted and readjusted and guessed at and written hastily because the time is up for a decision, etc., the derived quantum state should not be imagined to have any {\it ontic hold\/} on the world.  It is not that they are full of human error, but that they are full of humanity.  Quantum states are mental constructs whose building blocks are themselves subjective probability assignments, expectations, and gambling commitments.

\subsubsection{Interlude: Just Enough Quantum Information Theory}

This way of viewing the Born Rule can be made clearer and more general with the introduction of a little quantum information theory.  It is worthwhile going through this exercise to show off just how distinct QBism's way of viewing the Born Rule is.  At the exercise's end, the quantum state as a complex vector in Hilbert space and a measurement specification as a set of Hermitian operators will have completely disappeared from the formalism:  Only probabilities and conditional probabilities will remain, and the normative character of the Born Rule will be seen as a particular relation the agent should strive to maintain between his probability assignments.

With the advent of quantum information theory it was realized that the Dirac--von Neumann notion of a quantum measurement was too limiting for many natural problems.  Thus arose the notion of the positive-operator valued measure (or POVM) in the 1970s as the most general kind of quantum measurement conceivable~\cite{Nielsen10}.  In this theory, the outcomes or ``clicks'' $E_j$ of a measuring device are modeled by a {\it set\/} of positive semi-definite operators $\{\hat{E}_j\}$ on the Hilbert space ${\cal H}_d$ %\footnote{See Footnote \ref{FootNoteX} again.}
associated with the quantum system.\footnote{The $E_j$ with no hat denotes a lived experience, the click; the associated operator $\hat{E}_j$ is a tool for calculations.}
This means that the operators $\hat{E}_j$ are Hermitian with nonnegative eigenvalues and, together, fulfill the condition of being a resolution of the identity, i.e.,
\be
\sum_j \hat{E}_j = \hat{I}\;,
\ee
for the identity operator $\hat{I}$ on ${\cal H}_d$.  Beyond these conditions, however, there are no further restrictions on the operators $\hat{E}_j$.  For instance, even though the operators live on a $d$-dimensional Hilbert space, the index $j$ may take values from any range whatsoever, even a continuously infinite set.  Yet, in principle, any such set can made to correspond to the clicks of an appropriately constructed measuring device.

It is worth appreciating how different this notion of quantum measurement is from the notion found in Bohr's and the other founders' papers on quantum theory.  There, a measuring device was associated with a {\it single\/} Hermitian operator and the outcomes of the measurement with the {\it eigenvalues\/} of that operator.  Consequently, for a finite dimensional space ${\cal H}_d$, a quantum measurement could have at most $d$ distinct outcomes (taking into account that there might be degeneracies in the spectrum of the operator).  In the generalized notion of measurement, the outcomes are associated not with the eigenvalues of a single operator, but each with an operator itself.  If one wanted to think of the old notion of measurement in the new terms, then the outcomes $E_j$ of a measurement would be associated with the eigenprojectors $\hat{E}_j=|j\rangle\langle j|$ of the encompassing Hermitian operator, instead of its eigenvalues.

The most important thing this buys for the present purposes is that one can get a notion of {\it informational completeness\/} in the generalized setting just as in the preceding one, but now even with a {\it single\/} quantum measurement.  Informational completeness does not necessarily rely on the consideration of some number of complementary observables, as $\hat\sigma_x\,$, $\hat\sigma_y\,$, $\hat\sigma_z$ were.  In other words, one can now have a {\it single\/} quantum measurement through which the outcome probabilities uniquely specify  a $\hat\rho$.

The starting point is to look at how the Born Rule itself gets generalized for these generalized measurements.  The answer is, it doesn't.  For a quantum state $\hat\rho$, pure or mixed, and a measuring device described by a set of operators $\{\hat{E}_j\}$, the probabilities for the individual clicks $E_j$ are given by
\be
Q(E_j)= {\rm tr}\, \hat\rho\hat{E}_j \;.
\label{BR}
\ee
By the properties of the $\hat{E}_j$, one will have automatically that $Q(E_j)\ge0$ and $\sum_j Q(E_j) = 1$, exactly as required of a probability distribution, no matter what the density operator $\hat\rho$.  For a density operator itself is a positive semidefinite operator with trace 1.  Furthermore, in the special case that the measurement is one of the old von Neumann variety, Eq.\ (\ref{BR}) reduces to $Q(E_j)=\langle j|\hat\rho|j\rangle$. If we add to it that $\hat\rho$ is a pure state, i.e., $\hat\rho=|\psi\rangle\langle\psi|$, we get the further reduction to $Q(E_j)=|\langle\psi|j\rangle|^2$.  This much will be familiar to most readers.

Where the notion of informational completeness comes in can be found by looking at the right-hand side of Eq.\ (\ref{BR}).  The trace of the operator product of two Hermitian operators $\hat{A}$ and $\hat{B}$ is actually an inner product $(\hat{A},\hat{B})$---not an inner product on the space ${\cal H}_d$ itself, but on the linear vector space of Hermitian operators ${\cal L}({\cal H}_d)$.  So, the Born Rule boils down to nothing more than simply taking an inner product.  No fancy absolute-value-squared of an inner product, just an inner product itself.  What this means is that we may think of the right-hand side of Eq.\ (\ref{BR}) in simple geometrical terms:  It represents the projection of a vector $\hat\rho$ onto another vector $\hat{E}_j$.  Particularly, if we know the projections of a vector onto a complete basis for the vector space, then we can reconstruct the vector itself.

Putting these ideas together, if we can find any measuring device for which the $\{\hat E_j\}$ form a complete basis for the space ${\cal L}({\cal H}_d)$, then the probabilities $Q(E_j)$ will uniquely specify a quantum state $\hat\rho$.  Of course, not all measuring devices fulfill this condition:  For instance a measurement of the old von Neumann variety can't possibly fulfill it because there are only $d$ elements to it, and the dimensionality of the space ${\cal L}({\cal H}_d)$ is $d^2$.  It thus takes $d^2$ vectors to span the space.  Nonetheless, such informationally complete measurements abound---they exist in all dimensions~\cite{Caves02b}.  Once {\it any one such measurement\/} is singled out as being a special or {\it fiducial\/} measurement, one can rewrite all quantum states as the probability distributions they generate for that measurement's outcomes.  Consequently, one can say that a quantum state {\it just is\/} a probability distribution!  This is clearly the kind of language QBism feels the most comfortable speaking in.

But at what cost?  We can have a new formalism for quantum theory, one that involves probabilities only, but how ugly or mathematically recalcitrant might the theory become through this transformation?  For instance, what we have said so far establishes a one-to-one mapping between density operators and probability vectors,
\be
\hat\rho \quad \longleftrightarrow \quad \mathbf{Q} = \Big[ Q(E_1),\, Q(E_2),\, \ldots,\, Q(E_{d^2}) \Big]\;,
\label{Hermin}
\ee
so that we have a representation of quantum states living in a probability simplex rather than a complex vector space, but we have not established that the mapping is onto.  In fact, it is not, and regardless of the choice of our fiducial measurement it cannot be.  So instead of having the full probability simplex available for our $\mathbf{Q}$, there will be a potentially very complicated convex subset of the simplex in which the $\mathbf{Q}$ reside.  Philosophically, that is OK, because we wanted to imagine the Born Rule as a means of giving us guidance in the first place.  Part of that guidance might be, ``Hey that $\mathbf{Q}$ is a stupid choice; it doesn't live in the convex subset I was telling you about.''  Still, it does seem to indicate that one might want to very careful with the choice of the fiducial measurement so that, for instance, the allowed region for the $\mathbf{Q}$ becomes as mathematically simple as possible, or so that the representation becomes as filled with guidance as possible, or perhaps both.

This brings us to the cutting edge of QBism and also sets the stage for giving the mathematically simplest form by which to take the Born Rule as a normative relation.  For all quantum systems ${\cal H}_d$ with $d=2, 3, \ldots, 151$ (and a handful of other dimensions up to 844, so far), we know that there is a {\it very special\/} kind of informationally complete measurement available~\cite{Fuchs17,Scott17}---it goes by the name of a {\it symmetric informationally complete\/} (or SIC) measurement.  Let us call its outcomes $H_i$, $i=1,\ldots, d^2$.  As the name implies, the associated operators $\hat H_i$ can be chosen to have a certain (amazing) symmetry.  Each can be written in terms of a pure quantum state $\hat\Pi_i=|\psi_i\rangle\langle\psi_i|$ via
\be
\hat H_i = \frac{1}{d} \hat\Pi_i \;,
\ee
while pairwise the $\hat\Pi_i$ obtain completely uniform inner products with each other:
\be
{\rm tr}\, \hat\Pi_i\hat\Pi_k = \frac{1}{d+1} \qquad \mbox{for all } i\ne k\;.
\label{RubyNell}
\ee
Why call this symmetry amazing?  Well, each $\hat\Pi_i$ is specified by $2d-2$ real parameters, and there are $d^2$ of them, so we are looking to find $\approx 2d^3$ parameters to specify the full set.  On the other hand, there are $\frac{1}{2}(d^4 - d^2)\approx \frac{1}{2}d^4$ equations of constraint.  Of course, these are nonlinear equations, but at first sight the system might appear far too over-constrained to have any solutions at all.  Expressing a sentiment of William Wootters, one might wonder whether such structures even have a ``right to exist.''\footnote{In fact, in March 2003, a well-known mathematician from Germany wrote the author, ``It is only known for $n=2,3,8$ that there are $n^2$ equiangular complex lines, probably there are no more cases of existence.''}

It is a long story to tell why the SICs were first considered in QBism~\cite{Caves99}, but basically it had to do with early attempts to prove the quantum de Finetti theorem mentioned in the Introduction.  Since then, operator sets with this symmetry have become an object of study in their own right, with well more than 150 papers devoted to their further properties and structure~\cite{Fuchs17}. It is a very deep subject with mathematical tendrils running everywhere~\cite{Appleby15}.  Most recently the construction of these structures has even been found related to one of the remaining unsolved problems proposed by David Hilbert at the turn of the last century, Hilbert's 12th problem~\cite{Appleby17}.  If that counts as evidence of discovering a long-buried treasure within the structure of quantum theory, then so be it.  For our purposes, it is enough to know that most researchers in the quantum information community who work on this subject believe SICs not only exist in the dimensions mentioned, but in fact in {\it all\/} finite dimensions $d=2, 3, 4, \ldots$ (i.e., up to but not including $\infty$).  It just becomes a question of proving it (and reaping the spin-offs that might come from the effort).  {\it Meanwhile, let us suppose SICs do indeed exist in all finite dimensions.} We will show the pretty form the Born Rule takes under this supposition.

Let us consider again a {\it completely general\/} measurement $\{\hat D_j\}$, not just an informationally complete one as in the last five paragraphs\footnote{We call it $\{\hat D_j\}$ instead of $\{\hat E_j\}$ this time so as not to create any confusion with the previous discussion.}---it might be of the old von Neumann variety or still something else entirely. If an agent had assigned a quantum state $\hat\rho$, then just as before, the Born Rule would specify:
\be
Q(D_j)= {\rm tr}\, \hat\rho\hat{D}_j \;.
\label{Bijoux}
\ee
Now, for a SIC measurement, the operators $\hat H_i$ are used to calculate probabilities for its outcomes, but the projection operators $\hat\Pi_i$ from which the $\hat H_i$ are constructed can be viewed as pure quantum states in their own right.  Thus, supposing a quantum state $\hat\Pi_i$ had been assigned to the quantum system for whatever reason, let us ask:  What would the Born Rule tell us of the probability for an outcome $D_j$?  It is ${\rm tr}\, \hat\Pi_i \hat D_j$ of course, as usual, but let us codify this with a name,
\be
P(D_j|H_i) = {\rm tr}\, \hat\Pi_i \hat D_j\;.
\label{Punt}
\ee
We do this because the quantity can be viewed as a conditional probability distribution for $D_j$ conditioned on $H_i$ if we consider the right scenario.  For instance, one might imagine the agent starting with the quantum state $\hat\rho$, performing a SIC measurement on the system, and then passing it on to the general measurement $\{\hat D_j\}$.  In light of the intermediate SIC measurement, the agent should update from $\hat\rho$ to some new quantum state, and if he is using L\"uders rule \cite{busch2009luders} to do this, it would be to $\hat\Pi_i$.  So, if the agent happens to know which click $H_i$ occurred in such a {\it cascaded measurement}, but not which $D_j$, he would indeed write down $P(D_j|H_i)$ as the conditional probability for the latter measurement outcome.

But what about the probability of the first click in this scenario?  It too would be given by the Born Rule, but let us write it to look similar to Eq.~(\ref{Punt}):
\be
P(H_i) = {\rm tr}\, \hat\rho\hat{H}_i \;.
\label{Pass}
\ee
This is just the representation of $\hat\rho$ with respect to the SIC measurement, as in Eq.~(\ref{Hermin}), only using a different symbol this time around, namely $\mathbf{P} = \big[P(H_1), P(H_2), \ldots,\, P(H_{d^2})\big]$\@.  Furthermore, because of the special symmetry in Eq.~(\ref{RubyNell}), there is an impressively clean (and easy to work out) reconstruction of $\hat\rho$ in terms of the $\hat\Pi_i$:
\be
\hat\rho = \sum_{i=1}^{d^2}\left[(d+1)P(H_i)-\frac{1}{d}\right]\! \hat\Pi_i\;.
\label{LillyLilac}
\ee
Now, recall that $\mathbf{P}$ cannot be just any probability in the probability simplex, but must be confined to a convex region of a certain variety.  For instance, one easy-to-see property of the region is that for all the $\mathbf{P}$ in it, $P(H_i)\le \frac{1}{d}$ for all the $i$.  So, for no quantum state $\hat\rho$, can one ever be too sure of the outcome of a SIC measurement.  In the end, the precise conditions on $\mathbf{P}$ so that it is consistent with all aspects of the Born Rule is simply that the $\hat\rho$ reconstructed in Eq.~(\ref{LillyLilac}) be positive semi-definite.  In other words, $\mathbf{P}$ must be in a convex set whose extreme points satisfy the condition of being pure quantum states. These extreme points can be expressed in two equations: one a quadratic, ${\rm tr}\, \hat\rho^2=1$, which in the language of $\mathbf{P}$ becomes simply an equation for a sphere,
\be
\sum_i p(H_i)^2 = \frac{2}{d(d+1)}\; ;
\label{Lucy}
\ee
and one a cubic, ${\rm tr}\, \hat\rho^3=1$, which becomes,
\be
\sum_{ijk} c_{ijk}\, P(H_i)P(H_j)P(H_k)=\frac{d+7}{(d+1)^3}\;,
\label{Desi}
\ee
where the real numbers $c_{ijk}={\rm Re} \big[{\rm tr}\big(\hat\Pi_i\hat\Pi_j\hat\Pi_k\big)\big]$ have some very impressive symmetry properties of their own~\cite{Appleby15}.\footnote{An enthusiast for Wolfgang Pauli's psycho-physical speculations might be intrigued by the right-hand side of this strange equation, especially as it captures the key nontrivial detail about the shape of quantum-state space from a QBist perspective.  Implicit in it is the number 137!  See, for instance, Ref.~\cite{Enz83}:  ``This number 137 symbolized for Pauli the connection with the magic world of the alchemists which had much fascinated him. By a coincidence which might be called `cabbalistic', Wolfgang Pauli died in room 137 of the Red Cross Hospital in Zurich on 15 December 1958.'' I thank John M. Hawthorne III for this wonderful observation.}

We now have all the ingredients we need to write down the Born Rule as expressed in Eq.\ (\ref{BR}), but purely in terms of probabilities.  Simply plugging the representation (\ref{LillyLilac}) into Eq.~(\ref{Bijoux}) and collecting terms gives~\cite{Fuchs13a}:
\be
Q(D_j) = \sum_{i=1}^{d^2}\left[(d+1)P(H_i)-\frac{1}{d}\right]\! P(D_j|H_i)\;.
\label{TheOneRing}
\ee
And that's what we've been after.\footnote{Of course, this form supposes that a SIC always exists for any dimension $d$.  Otherwise, it is only good in 151 dimensions or so---a nasty tease to our theorizing!  But even if SICs don't always exist, we could still give a QBist-style explanation of the Born Rule by using one of those abundant (unsymmetrical) informationally complete POVMs mentioned above---i.e., ones that {\it are\/} always known to exist.  It is just that such a version of the Born Rule will not be so simple as this one.}

Notice that, perhaps contrary to expectation,
\be
Q(D_j) \ne \sum_{i=1}^{d^2}P(H_i) P(D_j|H_i)\;.
\label{Loser}
\ee
This discrepancy, or ones morally equivalent to it (as for instance Feynman's discussion of the double slit experiment), have often been advertised as a statement that probability theory itself must be modified when the concern is quantum phenomena~\cite{Feynman51}.  But QBism sees Eq.~(\ref{TheOneRing}) as strictly an {\it addition\/} to probability theory---nothing about standard probability theory is negated.  The reason there is no equality in Eq.~(\ref{Loser}) is because the pure $\{\hat D_j\}$-measurement and the cascaded measurement (i.e., first the $\{\hat H_i\}$ then the $\{\hat D_j\}$) refer to two different hypothetical scenarios, ones that cannot be performed simultaneously.  In that sense, they are {\it complementary}---or at least the scenario here shares this much with Bohr's own usage of the term ``complementary''~\cite[p.\ 57]{Bohr49-1}:
\bq\small
\noindent In fact, we must realize that in the problem in question we are not dealing with a {\it single\/} specified experimental arrangement, but are referring to {\it two\/} different, mutually exclusive arrangements.
\eq

One way to think of the fiducial SIC measurement is that it stands above all other quantum measurements---it is a single measurement that is complementary to all the others.  No matter what the $\{\hat D_j\}$, the $\{\hat D_j\}$ and $\{\hat H_i\}$ measurements cannot be made together.  This is ultimately the reason for Eq.~(\ref{TheOneRing}).  One might say that if a value for $\{\hat D_j\}$ is allowed to come into being, a value for $\{\hat H_i\}$ cannot be assumed sitting there unrevealed.  If it were otherwise, we would have equality in Eq.~(\ref{Loser}):  The $H_i$ would then be an ontic- or hidden-variable model for the $\{\hat D_j\}$ in the sense of Harrigan and Spekkens~\cite{Harrigan10}, and we would have used the standard Bayesian Law of Total Probability to calculate $Q(D_j)$.  Instead we only use the ingredients from that law, but in the new combination that is uniquely quantum---Eq.~(\ref{TheOneRing}).

As a normative relation, what the Born Rule Eq.~(\ref{TheOneRing}) is doing, is suggesting to the agent:  Before you gamble on the outcomes of this experiment, you should think hard about how you would gamble on the cascaded one as well. If you come up with sets of probabilities such that
\be
Q(D_j) \ne \sum_{i=1}^{d^2}\left[(d+1)P(H_i)-\frac{1}{d}\right]\! P(D_j|H_i)\;,
\ee
then you should strive harder to find a set of probability assignments which do instate equality.  You should dip back into your experience and reassess: This is the only suggestion of the Born Rule.  It doesn't single out any of the terms as more important than the others:  It might be $\mathbf{P}$ that the agent adjusts, or $\mathbf{Q}$, or some of the conditional probabilities, or some combination of all of the above.  It is the relation that is normative, not the terms within it.\footnote{This is one of the points of contrast between QBism and Richard Healey's  ``pragmatic interpretation of quantum mechanics''~\cite{Healey15}.}  As L. J. Savage says of probability theory~\cite{Savage54}
\bq\small
\noindent
According to the personalistic view, the role of the mathematical theory of probability is to enable the person using it to detect inconsistencies in his own real or envisaged behavior. It is also understood that, having detected an inconsistency, he will remove it. An inconsistency is typically removable in many different ways, among which the theory gives no guidance for choosing.
\eq
so too QBism says of the Born Rule.  Though here the inconsistency is not of a purely logical character, but between an agent's desire to survive and flourish in the world in which he is immersed and that world's actual character.

From this point of view, the Born Rule as a normative requirement binds the agent's gambles about the consequences of one hypothetical action on the world to the consequences of another hypothetical action.  But it may go still further:  There is a sense in which Eq.~(\ref{TheOneRing}) has so much Hilbert-space structure {\it already snuck into it}---in comparison to the analogous expressions for the Born Rule derived from other informationally complete measurements---that it may indeed bind quantum theory itself together.  This is to say, the consistency of this equation alone already implies significant features of Eqs.~(\ref{Lucy}) and (\ref{Desi}), and there is hope that with one well-motivated further requirement, it might just go all the way~\cite{Fuchs13a,Appleby09a,Appleby16b}.
%We called the SIC a kind of royal measurement for the simplicity it gives this normative requirement, but it really goes further than that.  If it weren't so sinister, one might think of Tolkien's One Ring in {\sl The Lord of the Rings\/}:
%\bv
%One Ring to rule them all, One Ring to find them,\\
%One Ring to bring them all and in the darkness bind them.
%\ev
%The SICs %, in the sense of giving the conceptually simplest representation of the Born Rule,
%may indeed bind all of quantum theory together~\cite{Fuchs13a,Appleby09a,Appleby16b}, but in this case it is a wonderful, miraculous thing, not sinister.

Finally, there are two special cases of Eq.~(\ref{TheOneRing}) worth noting.  The first concerns the case where the $\hat D_j=|j\rangle\langle j|$ correspond to a von Neumann measurement.  Then our normative relation reduces to a simple deformation of the Law of Total Probability:
\be
Q(D_j) = (d+1) \sum_{i=1}^{d^2} P(H_i)P(D_j|H_i)\, - \,1\;.
\ee
This case helps isolate the extra role of the dimension $d$ in the Born Rule in comparison to the Law of Total Probability:  It is a measure of the deviation of the former from the latter.  Thus, as dimension grows, it is not that we return to an unaugmented ``classical probability theory'' in our thinking about quantum systems, but ever further deviate from it.

In a second case, this functional form appears again, though with a different constant at the tail end.  Suppose $\{\hat D_j\} = \{\hat U^\dagger \hat H_j \hat U\}$ is a unitarily rotated version of the fiducial SIC.  This, for instance, can be thought of as a time-evolved SIC in the Heisenberg picture of quantum dynamics.  Since,
\be
Q(D_j) = {\rm tr}\! \left[\hat \rho\, (\hat U^\dagger \hat H_j \hat U)\right] = {\rm tr}\! \left[(\hat U \hat \rho \hat U^\dagger)\, \hat H_j\right]
\ee
we might relabel this probability distribution as $P_t (H_j) = Q(D_j)$ to help emphasize that if $P(H_i)$ is the SIC representation of $\hat\rho$, then $P_t (H_j)$ is the SIC representation of  $\hat U \hat \rho \hat U^\dagger$, the time-evolved state.  In this case now, Eq.~(\ref{TheOneRing}) reduces to
\be
P_t (H_j) = (d+1) \sum_{i=1}^{d^2} P(H_i)P(D_j|H_i)\, - \,\frac{1}{d}\;.
\ee
Particularly note that, from this point of view, even the Schr\"odinger equation becomes a normative relation!  For, the content of this equation {\it just is\/} the Schr\"odinger equation; it is merely another species of the Born Rule---one about how an agent should gamble {\it now\/} if he thinks he would instead gamble some other way {\it later}, or vice versa.

Never mind what Bohr might have meant by ``the quantum postulate,''\footnote{See page \pageref{QuantPost} again.} if anything stands out as a candidate {\it quantum postulate\/} for QBism, it is Eq.~(\ref{TheOneRing})~\cite{Fuchs13a,Appleby16b}.

\subsection{Second Tenet: My Probabilities Cannot Tell Nature What To Do}
\label{TenetTwo}

This leads to a second tenet of QBism.  A woman believes with all her heart and soul that she will never find her husband cheating on her.  Yet no matter how strong her belief, we know that her husband need not comply to it---tragedy sometimes happens.  The woman's beliefs or gambling commitments are one thing; the events of the world are another.  This is what QBists mean by saying that a quantum state has no ``ontic hold'' on the world.  An agent's mesh of beliefs may cause her to make a pure quantum-state assignment for some quantum system, and she could even consider a yes-no measurement that is exactly the projection onto that state.  Using the Born Rule, the agent would then calculate the probability of a ``yes'' outcome to be {\it exactly\/} 1---i.e., she should believe a ``yes'' will occur with all her heart and soul.  But that does not mean that the world must comply.  The agent's quantum state assignment does not mean that the world is {\it forbidden\/} to give her a ``no'' outcome for this measurement~\cite{Caves07}.  All {\it she\/} knows by using the Born Rule is that {\it she\/} has made the best gamble {\it she\/} could in light of all the other gambling commitments {\it she\/} might be making with regard to other phenomena (other experiments, etc.).

There is a sense in which this unhinging of the Born Rule from being a ``law of nature'' in the usual conception~\cite[pp.~86--101]{Weinberg15}---i.e., treating it as a normative statement, rather than a descriptive one---makes the QBist notion of quantum {\it indeterminism\/} a far more radical variety than anything proposed in the quantum debate before.  It says in the end that nature does what it wants, without a mechanism underneath, and without any ``hidden hand'' \cite{Fine89} of the likes of Richard von Mises's {\it Kollective\/} or Karl Popper's {\it propensities\/} or David Lewis's {\it objective chances\/}, or indeed any conception that would diminish the autonomy of nature's events.\footnote{This idea may bear a resemblance in certain respects to the concept of ``Knightian freedom'' introduced by Scott Aaronson~\cite{Aaronson13}, but we leave this question for future research.}  Nature and its parts do what they want, and we as free-willed agents do what we can get away with.  Quantum theory, on this account, is our best means yet for hitching a ride with the universe's pervasive creativity and doing what we can to contribute to it.

Einstein put it nicely once in a letter to Fritz Reich and his wife~\cite[p.~390]{Stachel02},
\bq\small
\noindent I still do not believe that the Lord God plays dice. If he had wanted to do this, then he would have done it quite thoroughly and not stopped with a plan for his gambling: In for a penny, in for a pound [Wenn schon, denn schon].  Then we wouldn't have to search for laws at all.
\eq
My colleague R\"udiger Schack put the appropriate response even more aptly, ``God {\it has\/} done it quite thoroughly.  That's the message of QBism.  It is not a plan for {\it his\/} gambling, but for {\it ours}.''

\subsection{Third Tenet: A Measuring Device Is Literally an Extension of the Agent}

The first two tenets of QBism already spook a lot of people, but the third one often takes the prize for the most frightening.  It's enough to cause some to run away screaming {\it Solipsism!} (Luckily a few have come back, but I suspect some never will~\cite{Norsen16,Gisin17}\@.)  So, get ready.

In the discussion of the second tenet, I deliberately made it hard to not notice how often the word {\it she\/} was emphasized.  This was given a second-person perspective for presentational purposes, but the message was intended to be about the first-person.  This is because QBism's understanding of quantum theory is purely in first-person terms.  When I---the agent---write down a quantum state, it is {\it my\/} quantum state, no one else's.  When I contemplate a measurement, I contemplate its results, outcomes, consequences {\it for me}, no one else---it is {\it my\/} experience.  This is why it was declared that any measuring equipment should literally be considered part of the agent.  But of course, it is the same for {\it you\/} when {\it you\/} are the one applying quantum theory.  It is what it means for quantum theory to be a ``user's manual'' or handbook for each of us~\cite{Varenna17}.

The roots of this idea come from Heisenberg and Pauli but go much further.  In a 1935 paper titled, ``Questions of Principle in Modern Physics,'' Heisenberg wrote~\cite{Heisenberg35},
\bq\small
\noindent
In this situation it follows automatically that, in a mathematical treatment of the process, a dividing line must be drawn between, on the one hand, the apparatus which we use as an aid in putting the question and thus, in a way, treat as part of ourselves, and on the other hand, the physical systems we wish to investigate.
\eq
which Pauli seemed to endorse and take still further.  For instance, in a 1955 letter to Bohr, Pauli wrote~\cite{Pauli55},
\bq\small
\noindent
As it is allowed to consider the instruments of observation as a kind of prolongation of the sense organs of the observer, I consider the impredictable change of the state by a single observation---in spite of the objective character of the result of every observation and notwithstanding the statistical laws for the frequencies of repeated observation under
equal conditions---to be {\it an abandonment of the idea of the isolation (detachment) of the observer from the course of physical events outside himself}.
\eq
So the idea was certainly in the air with at least some of the Copenhageners, but one has to wonder how seriously they took themselves.  By 1958 Pauli seemed to be on the retreat~\cite{Pauli58a}:
\bq\small
The objectivity of physics is however fully ensured in quantum
mechanics in the following sense.  Although in principle, according to
the theory, it is in general only the statistics of series of
experiments that is determined by laws, the observer is unable, even
in the unpredictable single case, to influence the result of his
observation---as for example the response of a counter at a particular
instant of time.  {\it Further, personal qualities of the observer
do not come into the theory in any way---the observation can be made by
objective registering apparatus, the results of which are objectively
available for anyone's inspection.}  [my emphasis]
\eq

QBism, in contrast, takes the idea of ``the instruments of observation as a \ldots\ prolongation of the sense organs of the observer'' deadly seriously and runs it to its logical conclusion.  This is why QBists opt to say that the {\it outcome\/} of a quantum measurement is a {\it personal experience\/} for the agent gambling upon it.  Whereas Bohr always had his classically describable measuring devices mediating between the registration of a measurement's outcome and the individual agent's experience, for QBism the outcome {\it just is\/} the experience.  This is the aspect of QBism that takes it firmly into the territory of early American pragmatism~\cite{James96a,James96b}.

The reason for this move has to do with the desire to maintain quantum theory as a universally valid schema:  a handbook that anyone can use, in principle, for better gambling on the consequences of any action on {\it any\/} physical system.  This forces a reconsideration of the Wigner's friend paradox, a place many Copenhageners don't seem to want to go~\cite{Englert13}.  The key realization is that this length is required to save the consistency of QBism:  When Wigner writes down a quantum state for {\it any\/} quantum system---say, the compound system of an electron plus his friend, who himself is believed to be performing a spin-measurement on the electron---Wigner is {\it never\/} telling a story~\cite[p.\ 592]{Timpson08}\footnote{Furthermore, this is one of the essential ingredients in how QBism evades the recent very strong no-go theorem of Frauchiger and Renner~\cite{Renner16}.  In fact, QBists are quite grateful for their introducing the concept of ``physical theories as constraints on stories''---a notion which captures all other interpretations of quantum theory {\it except\/} QBism.  From our point of view, their theorem should be interpreted as a rather nice piece of theoretical evidence in favor of QBism: For it aims to show that if one desires an interpretation falling within their framework, it had better be Everettian quantum mechanics or else.  See also Ref.~\cite{Brukner17} for a related thought experiment, but one which uses it, in contrast to Frauchiger and Renner, as an argument for a quantum interpretation to some extent aligned with QBism.} of what is happening on the inside of the conceptual box surrounding these things~\cite[Section III]{Fuchs10a}\@.  Nor is the unitary evolution he ascribes to the conceptual box telling a story of the actual goings-on within it.  Wigner's quantum-state assignment and unitary evolution for the compound system are only about his {\it own\/} expectations for his {\it own\/} experiences should he take one or another action upon the system or any part of it.  One such action might be his sounding the verbal question, ``Hey friend, what did you see?'', which will lead to one of two possible experiences for him.  Another such action could be to put the whole conceptual box into some amazing quantum interference experiment, which would lead to one of two completely different experiences for him.  The friend on the inside, of course, has his own story to tell, but that's his business, not Wigner's.  In fact, the closest QBism comes to telling any story at all is to say, whatever is in the conceptual box---whatever is happening in there---Wigner would be well-advised to adopt the Born Rule for gambling upon its potential impact to {\it him}.

Even though it concerns a different scenario, it is useful to compare this to David Wallace's very clear description of a Bell experiment from the Everettian point of view~\cite[p.~310]{Wallace12}:
\bq\small
[I]n Everettian quantum mechanics, violations of Bell's inequality are relatively uninteresting.  For Bell's theorem \ldots\ simply does not apply to the Everett intepretation. It assumes, tacitly, among its premises that experiments have unique, definite outcomes.

From the perspective of a given experimenter, of course, her experiment {\it does\/} have a unique, definite outcome, even in the Everett interpretation. But Bell's theorem requires more: it requires that from her perspective, her distant colleague's experiment also has a definite outcome.  This is not the case in Everettian quantum mechanics---not, at any rate, until that distant experiment enters her past light cone. And from the third-person perspective from which Bell's theorem is normally discussed, no experiment has any unique definite outcome at all.
\eq
For QBism, the much confirmed experimental violations \cite{Hensen15,Giustina15,Shalm15} of Bell's inequality---confirmed like beating a dead horse---are not at all uninteresting~\cite[Section V]{Fuchs10a}\@. This is because these observations demand a serious reconsideration of the kind of realism, {\it participatory realism}~\cite{Fuchs16}, QBism is an example of---a variety of realism far too often left out of philosophy-of-science discussions.  (Though see Ref.~\cite{Ben-Menahem97} for a potentially sympathetic view.)

On the other hand, this quote clearly shows an element in common between the Everettian analysis of Bell and the QBist analysis of Wigner's friend.  ``For a given experimenter, her experiment {\it does\/} have a unique, definite outcome''---this is true in both Everett and QBism.  ``Bell's theorem requires more''---once again, true for both interpretations~\cite{Fuchs14}\@.  But what of ``the distant colleague''?  In QBism, it is a physical system like any other in the world external to the ``given experimenter.''  It is thus in the ``given experimenter's'' interest to use the methods of quantum theory to analyze that system, rather than base her analysis on some supposed {\it hidden variables\/} over there---namely, the ``distant colleague's definite outcomes.''  (For, what else would they be but hidden variables to the ``given experimenter''?)

The question now arises, is this different in any serious way from the Everettian's ``[that the] distant colleague's experiment has a definite outcome \ldots\ is not the case in Everett\ldots''?  Well, it is.  From the QBist perspective, the Everettian statement goes too far:  As \cite{Renner16} makes clear with newfound precision, it tells a story of what is and is not going on inside the conceptual box.  As QBism eschews this in the case of Wigner's friend, it also eschews it in the case of the Bell scenario.  At the root of this is one of the most distinguished differences between Everett and QBism.  \textit{``QBism don't do third-person!''}  For QBism, all of quantum theory is first-person for the person who happens to be using it.

So there we have it, three tenets of QBism:  The Born Rule, the very foundation of what quantum theory means for QBism, is {\it normative}.  Probabilities are so subjective, they {\it never tell\/} the world what to do.  And quantum measurement outcomes {\it just are\/} personal experiences.  Now it is time to talk about Bohr in more detail.

\section{Notwithstanding Bohr}

\subsection{Pauli and Bohr on the Detached Observer}
\label{PauliBohrDebate}

\medskip

\begin{flushright}
\baselineskip=13pt
\parbox{4.0in}{\baselineskip=13pt\footnotesize
Notwithstanding all differences, a certain analogy between the postulate of relativity and the point of view of complementarity can be seen in this, that according to the former the laws which in consequence of the finite velocity of light appear in different forms depending on the choice of the frame of reference, are equivalent to one another, whereas, according to the latter the results {\bf obtained by different measuring arrangements} apparently contradictory because of the finite size of the quantum of action, are logically compatible.}\\
\footnotesize --- Niels Bohr \cite[p.\ 85]{Bohr37-2}
\end{flushright}

Imagine two grown men fighting over who possesses the meaning of a two-word phrase; seems silly, doesn't it?  But when those two men are Bohr and Pauli, you can guess something big was at stake.  The trouble was over the notion of a ``detached observer.''  The disagreement started in a letter from Pauli to Bohr dated 15 February 1955, where Pauli writes,\footnote{I first encountered this correspondence in \cite{Laurikainen88} (first published in Finnish in 1985), and then later again in \cite{Folse85} (also published in 1985).  A very enlightening discussion on the issues at stake can be found in \cite[pp.\ 131--135]{Gieser05}.  I thank Henry Folse for supplying me with copies of the original correspondence in full.}
\bq\small
It is with great pleasure that I received your nice letter and above all, the text of your lecture on ``Unity of Knowledge''. The general outlook of it is of course the same as mine.  Under your great influence it was indeed getting more and more difficult for me to find something on which I have a different opinion than you. To a certain extent I am therefore glad, that eventually I found something: the definition and the use of the expression ``detached observer'' \ldots. According to my own point of view the degree of this ``detachment'' is gradually lessened in our theoretical explanation of nature and I am expecting further steps in this direction.
\eq
It seems worthwhile to quote this letter at length to help set the stage for where QBism picks up in the argument.\footnote{In the correspondence between Pauli and Bohr, their letters are in the original---i.e., these are not translations from German or Danish, but were originally composed in English. The spelling errors and grammatical mistakes are taken verbatim from the originals.}

\bq\small
[I]t seems to me quite appropriate
to call the conceptual description of nature in classical physics,
which Einstein so emphatically wishes to retain, ``the ideal of the
detached observer''. To put it drastically the observer has
according to this ideal to disappear entirely in a discrete manner
as hidden spectator, never as actor, nature being left alone in a
predetermined course of events, independent of the way in which the
phenomena are observed. ``Like the moon has a definite position''
Einstein said to me last winter, ``whether or not we look at the
moon, the same must also hold for the atomic objects, as there is no
sharp distinction possible between these and macroscopic objects.
Observation cannot {\it create\/} an element of reality like a
position, there must be something contained in the complete
description of physical reality which corresponds to the {\it
possibility\/} of observing a position, already before the
observation has been actually made.'' I hope, that I quoted Einstein
correctly; it is always difficult to quote somebody out of memory
with whom one does not agree. It is precisely this kind of postulate
which I call the ideal of the detached observer.

In quantum mechanics, on the contrary, an observation hic et nunc
changes in general the ``state'' of the observed system in a way not
contained in the mathematically formulated {\it laws}, which only
apply to the automatical time dependence of the state of a {\it
closed\/} system. I think here on the passage to a new phenomenon by
observation which is technically taken into account by the so called
``reduction of the wave packets.'' As it is allowed to consider the
instruments of observation as a kind of prolongation of the sense
organs of the observer, I consider the impredictable change of the
state by a single observation---in spite of the objective character
of the result of every observation and notwithstanding the
statistical laws for the frequencies of repeated observation under
equal conditions---to be {\it an abandonment of the idea of the
isolation (detachment) of the observer from the course of physical
events outside himself}.

To put it in nontechnical common language one can compare the role of
the observer in quantum theory with that of a person, who by its
freely chosen experimental arrangements and recordings brings forth a
considerable ``trouble'' in nature, without being able to influence
its unpredictable outcome and results which afterwards can be
objectively checked by everyone.

Probably you mean by ``our position as detached observers'' something
entirely different than I do, as for me this new relation of the
observer to the course of physical events is entirely {\it
identical\/} with the fact, that our situation as regards objective
description in ``this field of experience'' gave rise to the demand
of a renewed revision of the foundation for ``the unambiguous use of
our elementary concepts'', logically expressed by the notion of
complementarity.
\eq
The contrast with Bohr is immediate. In a 2 March 1955 reply, Bohr writes,
\bq\small
\noindent Of course,
one may say that the trend of modern physics is the attention to the
observational problem and that just in this respect a way is bridged
between physics and other fields of human knowledge and interest. But
it appears that what we have really learned in physics is how to
eliminate subjective elements in the account of experience, and it
is rather this recognition which in turn offers guidance as regards
objective description in other fields of science. To my mind, this
situation is well described by the phrase `detached observer', and it
seems to me that your reference to our controversy with Einstein is
hardly relevant in this connection. Just as Einstein himself has
shown how in relativity theory `the ideal of the detached observer'
may be retained by emphasizing that coincidences of events are common
to all observers, we have in quantum physics attained the same goal
by recognizing that we are always speaking of well defined
observations obtained under specified experimental conditions. These
conditions can be communicated to everyone who also can convince
himself of the factual character of the observations by looking on
the permanent marks on the photographic plates. In this respect, it
makes no difference that in quantum physics the relationship between
the experimental conditions and the observations are of a more
general type than in classical physics. I take it for granted that,
as regards the fundamental physical problems which fall within the
scope of the present quantum mechanical formalism, we have the same
view, but I am afraid that we sometimes use a different terminology.
Thus, when speaking of the physical interpretation of the formalism,
I consider such details of procedure like `reduction of the wave
packets' as integral parts of a consistent scheme conforming with
the indivisibility of the phenomenon and the essential
irreversibility involved in the very concept of observation. As
stressed in the article, it is also in my view very essential that
the formalism allows of well defined applications only to closed
phenomena, and that in particular the statistical description just
in this sense appears as a rational generalization of the strictly
deterministic description of classical physics.
\eq
There followed two more exchanges between the men and at least one face-to-face conversation. However, Pauli stated in a 1956 letter to Franz Kr\"oner~\cite[p.\ 133]{Gieser05} that the ``mishap'' over their differing uses of the term ``detached observer'' never reached a satisfactory conclusion.

What stance does QBism take on the matter?  Well, it certainly sides with Pauli over Bohr that ``the degree of this `detachment' is gradually lessened in our theoretical explanation of nature.''   But when Pauli hints, ``I am expecting further steps in this direction,'' QBism says disappointedly, ``Yeah, you really didn't take it far enough.''

The chain of reasoning goes like this.  When Bohr writes, ``Just as Einstein himself has shown how in relativity theory `the ideal of the detached observer' may be retained by emphasizing that coincidences of events are common to all observers, we have in quantum physics attained the same goal by recognizing that we are always speaking of well defined observations obtained under specified experimental conditions. These conditions can be communicated to everyone who also can convince himself of the factual character of the observations by looking on the permanent marks on the photographic plates,'' QBism asks, {\it ``But then what about Wigner's friend?''}  On the other hand, when Pauli comes along and says, ``As it is allowed to consider the instruments of observation as a kind of prolongation of the sense organs of the observer, I consider the impredictable change of [nature]\footnote{With perhaps consequential modification on our part with the insertion of ``nature'' in place of his ``state'' at this point.} by a single observation--- \ldots\ ---to be {\it an abandonment of the idea of the isolation (detachment) of the observer from the course of physical events outside himself},'' QBism says, {\it ``Ah, now we're on the right track!  This is exactly what Tenet 3 is all about.''}  \ldots\ But maybe we speak too soon. For when Pauli fills in the ellipsis with ``in spite of the objective character of the result of every observation and notwithstanding the statistical laws for the frequencies of repeated observation under equal conditions,'' QBism asks, {\it ``Again, what about Wigner's friend?''}  It's one step forward, two steps back.

The issue is always the same.  When Pauli, in his follow-on letter of 11 March 1955, writes,
\bq\small
\noindent As is well known to both of us, it is essential in quantum mechanics that the apparatus can be described by classical concepts. Therefore the observer is always entirely detached to the {\it results\/} of his observations (marks on photographic plates etc.), just as he is in classical physics.
\eq
or Bohr, in his Rutherford Memorial Lecture~\cite[p.\ 59]{Bohr61-1} adds to it,
\bq\small
\noindent In order that the inquiry may augment common knowledge, it is an obvious demand that the recording of observations as well as the construction and handling of the apparatus, necessary for the definition of the experimental conditions, be described in plain language. In actual physical research, this demand is amply satisfied with the specification of the experimental arrangement through the use of bodies like diaphragms and photographic plates, so large and heavy that their manipulation can be accounted for in terms of classical physics, although of course the properties of the materials of which the instruments, as well as our own bodies, are built up depend essentially on the constitution and stability of the component atomic systems defying such account.
\eq
they shun the demand that an analysis of Wigner's friend should be given in purely quantum theoretical terms.  In fact they promulgate the opposite---that in order to make sense of the practice of quantum theory, there must be a place in one's analysis where one can conveniently revert to a description of external physical objects (particularly, ``the measuring devices'') in classical terms.  The formalism of quantum theory for them thus becomes a less-than-universal tool, always needing a supplement drawn somewhere from a classical worldview.  For, who are Wigner and his friend in their eyes but characters in the story ``that the results of observation, which can be checked by anyone, cannot be influenced by the observer, once he has chosen his experimental arrangement''~\cite{Pauli58b}.  Asher Peres would say, ``Give me enough money and I will help Wigner manipulate his friend so meticulously that he'll not be able to answer what he saw on the photographic plate, but he will be able to tell me which superposition he felt in the Fourier basis.''

Leaving aside this QBistic criticism of Bohr's and Pauli's takes on the scope of the quantum formalism, there is nonetheless something in this correspondence that QBism would do well to think upon for further inspiration.  For, remember, the role of the hero is not that his words should be written in stone, but that they should help set the agenda and constellate one's thoughts.  In the same 11 March 1955 follow-on letter quoted above, Pauli lays out what seems to be a prescient program for QBism's concerns,
\bq\small
As I was mostly interested in the question, {\it how much informative reference to the observer an objective description contains}, I am emphasizing that a communication contains in general {\it informations on the observing subject}.

Without particularly discussing the separation between a subject and the informations about subjects (given by themself or by other persons), which can occur as elements of an ``objective
description'', I introduced a concept ``degree of detachment of the observer'' in a scientific theory to be judged on the kind and measure of informative reference to the observer, which this description contains.  For the objective character of this description it is of course sufficient, that every individual observer can be replaced by every other one which fulfills the same conditions and obeys the same rules.  In this sense I call a reference to experimental conditions an ``information on the observer'' (though an impersonal one), and the establishment of an experimental arrangement fulfilling specified conditions an ``action of the observer''---of course not of an individual observer but of ``the observer'' in general.

In physics I speak of a detached observer in a general conceptual description or explanation only then, {\it if it does not contain any explicit reference to the actions or the knowledge of the observer}.  The ideal, that this should be so, I call now ``the ideal (E)'' in honor of Einstein.  Historically it has its origin in celestial mechanics.
\eq

Note first that Pauli calls the ``establishment of an experimental arrangement'' an ``action of the observer.''  That certainly anticipates QBism's Tenet 3 and is quite likely a part of the psychological undercurrent leading to QBism's calling a quantum measurement an action upon a system.\footnote{Though the terminology in Leonard J. Savage's development of Bayesian decision theory \cite{Savage54}, with its ``decisions,'' ``acts,'' and ``consequences,'' certainly played a role there too.}  But what is of more interest for the present concerns of Tenet 1---i.e., trying to find a good representation of quantum theory that underlines the normative character of the Born Rule---is Pauli's notion of a ``degree of detachment.''

``For the objective character of this description it is of course sufficient, that every individual observer can be replaced by every other one which fulfills the same conditions and obeys the same rules,'' Pauli writes.  In QBism, the only piece of the quantum formalism that plays an objective role is the normative character of the Born Rule.  The Born Rule is a relation that {\it any agent\/} should strive to attain in his probabilistic assignments, so that he might become a ``better bettabilitarian.''  The elements within any application of the Born Rule are subjective, but the rule itself is objective:  ``Every individual observer can be replaced by every other one which \ldots\ obeys the same rules.''  (See Ref.~\cite{Fuchs16} for more details.)

So, what might play the role of a degree of detachment within QBism? In Eq.~(\ref{TheOneRing}), we saw that the Born Rule can be written as
\be
Q(D_j) = \sum_{i=1}^{d^2}\left[(d+1)P(H_i)-\frac{1}{d}\right]\! P(D_j|H_i)\;.
\label{Rewind}
\ee
In opposition to quantum theory, if the agent had instead been completely detached as in the ``ideal (E),'' so as not to need take account of the two complementary scenarios in any essential way, he would have written instead the more conventional Bayesian formula \cite{Fuchs13a},
\be
Q(D_j) = \sum_{i=1}^{d^2}P(H_i) P(D_j|H_i)\;.
\label{Loser2}
\ee
This suggests that the factors $(d+1)$ and $\frac{1}{d}$ in Eq.~(\ref{Rewind}) may be connected intimately to the degree to which QBism sees the agent engaged in the phenomena his actions help bring about.  Indeed one can formulate quantum theory over real number fields and quaternionic number fields, instead of the usual complex field, and one finds in those dimensions where the Born Rule is derivable under similar assumptions as Eq.~(\ref{Rewind}), its form becomes the same, but with different factors in place of the $(d+1)$ and $\frac{1}{d}$.  For instance for real vector spaces, the factors become $\big(\frac{1}{2}d+1\big)$ and $\frac{1}{d+1}$, respectively.  More generally, one can consider theories where the Born Rule takes the form
\begin{equation}
Q(D_j)=\sum_{i=1}^n \Big[\alpha\/ P(H_i) - \beta\Big] P(D_j|H_i)\;,
\end{equation}
for real numbers $\alpha$ and $\beta$, and try to find what makes quantum theory special along this spectrum.  This, however, is an ongoing research project---an ongoing research project stimulated by the Bohr-Pauli debate over a two-word phrase!~\cite{Varenna17,Fuchs11b}

\subsection{Bohr on EPR, QBism on EPR}

\medskip

\begin{flushright}
\baselineskip=13pt
\parbox{4.0in}{\baselineskip=13pt\footnotesize
Notwithstanding the power of quantum mechanics as a means of ordering an immense amount of evidence regarding atomic phenomena, its departure from accustomed demands of causal explanation has naturally given rise to the question {\bf whether we are here concerned with an exhaustive description of experience}.}\\
\footnotesize --- Niels Bohr \cite[p.\ 3]{Bohr58-1}
\end{flushright}

The Einstein, Podolsky, Rosen paper of 1935 \cite{EPR35} is currently cited over 16,000 times on Google Scholar.  That's nearly 4,000 more times than the original quantum teleportation paper \cite{Bennett93}, which in the past I have called the beginning of the field of quantum information~\cite[p.\ xxv]{Fuchs10}.  So the EPR paper must be important!  %\smiley

Anyone reading this essay will know the EPR paper intimately, as well as its famous ``criterion of reality'':
\bq\small
\noindent If, without in any way disturbing a system, we can predict with certainty (i.e., with probability equal to unity) the value of a physical quantity, then there exists an element of reality corresponding to that quantity.
\eq
Of course, EPR wanted to use this to show that quantum theory ought to be considered incomplete.  Perhaps a more direct statement of what Einstein was really after can be found in a letter to Karl Popper dated 11 September 1935~\cite{Popper02}:
\bq\small
It should be noticed that some of the precise predictions which I can
obtain for the system B (according to the freely chosen way of measuring
A) may well be related to each other in the same way as are
measurements of momentum and of position. One can therefore
hardly avoid the conclusion that the system B has indeed a definite
momentum and a definite position co-ordinate. For if, upon freely
choosing to do so [that is, without interfering with it],\footnote{This bracketed comment is presumably Popper's own.} I am able to
predict something, then this something must exist in reality.
\eq

For Bohr, the EPR paper was another opportunity to rethink the lessons of complementarity~\cite{Bohr35}:
\bq\small
From our point of view we now see that the wording of the above-mentioned criterion of physical reality proposed by Einstein, Podolsky and Rosen contains an ambiguity as regards the meaning of the expression ``without in any way disturbing a system.'' Of course there is in a case like that just considered no question of a mechanical disturbance of the system under investigation during the last critical stage of the measuring procedure. But even at this stage there is essentially the question of {\it an influence on the very conditions which define the possible types of predictions regarding the future behavior of the system}. Since these conditions constitute an inherent element of the description of any phenomenon to which the term ``physical reality'' can be properly attached, we see that the argumentation of the mentioned authors does not justify their conclusion that quantum-mechanical description is essentially incomplete. On the contrary this description, as appears from the preceding discussion, may be characterized as a rational utilization of all possibilities, of unambiguous interpretation of measurements, compatible with the finite and uncontrollable interaction between the object and the measuring instruments in the field of quantum theory. In fact, it is only the mutual exclusion of any two experimental procedures, permitting the unambiguous definition of complementary physical quantities, which provides room for new physical laws the coexistence of which might at first sight appear irreconcilable with the basic principles of science. It is just this entirely new situation as regards the description of physical phenomena that the notion of complementarity aims at characterizing.
\eq
To this day, it remains hard for me to parse what Bohr was saying in this reply.  But I have been able to gather this much: He believes the EPR reality criterion contains an ambiguity in the meaning of ``if without in any way disturbing a system.''  That much is plain.  So at the least, we must ask how does this relate to QBism's response to EPR?

QBism too thinks there is an essential ambiguity in the reality criterion, but it has less---if anything at all---to do with the ``if without in any way disturbing a system'' clause than Bohr seems to think.  This much is further clear about Bohr: {\it Whatever he is saying}, it seems to have essentially nothing to do with the modern commentaries on Bell and EPR that are rife at the philosophy-of-physics meetings where these things are usually discussed~\cite{Norsen11,Albert09}.  In that community, it is a tacit assumption that at least one part of the EPR criterion is unquestionable. Perhaps this was put most forcefully by Tim Maudlin~\cite{Maudlin14}:
\bq\small
If I could make one change to the EPR paper in retrospect it would be to alter the characterization of this criterion. The authors call it ``reasonable'' and ``in agreement with classical as well as quantum-mechanical ideas of reality'', but its status is actually much stronger than that: the criterion is, in the parlance of philosophers, {\it analytic}. That is, this criterion follows just from the very meanings of the words used in it. The difference is this: one {\it can\/} coherently (but not reasonably!)\ deny a merely reasonable claim, but one can't coherently deny an analytic proposition. Some people, hell-bent on denying the conclusion of the EPR argument, have taken to thinking they just have to reject this EPR criterion. If this can be done, then the conclusion need not be accepted.
\eq
The kind of narrative then given to accompany this is the following: The EPR argument establishes that {\it if there is no disturbance}, then there must be local elements of reality associated with quantum systems.  (They love to point out that local hidden variables are not an assumption, but a conclusion drawn from the EPR argument alone.)  But the Bell inequality violations exhibited in so many careful experiments establish that there are no such local hidden variables.  Consequently, it must be the case that the disturbance clause is the part of the criterion that is violated by our world.  \textit{Voila!}  Nature is nonlocal; there actually is spooky action at a distance! It does not dawn on Maudlin and like-minded philosophers of science that there might be another option: After all the EPR criterion is analytic, {\it right?}  And there is only one clause in it: Namely, {\it if there is no disturbance}.

Bohr was surely saying none of that.  Indeed, for all I know, what he did say, if only understood properly, might be something QBism would endorse.  We might look back and say, ``Ah, that's what Bohr was saying!  It took us {\it so long\/} to get to the same place.  If we had only understood him from the beginning.''  But, here it seems to me is an example where Ed Jaynes was dead on the mark with his quantum omelette.  For Bohr seems to be speaking of some kind of epistemic notion when he declares an ambiguity in the ``without in any way disturbing a system'' clause.  But who without a developed palate could tell?  Historical confusions abound.  On the other hand, physics as a practice should move past strained formulations when something clearer and more direct comes along.  This is what QBism attempts to supply.

So, how does QBism respond to the EPR argument?  With something so simple as this:  ``Tenet 2: My probabilities cannot tell nature what to do.''  Einstein, in his streamlining of the EPR exposition wrote, ``if, upon freely choosing to [measure something on A without interfering with B], I am able to predict something [on B], then this something must exist in reality.''  Predict, predict, what does it mean to predict?\footnote{Cf.\ Wheeler on Bohr, ``To be, to be, what does it mean to be?'' \cite[p.\ 131]{Wheeler98}}  To the subjective Bayesian, it means how an agent will gamble.  And in the quantum case, it means to the QBist how he will gamble on his own fresh experience, as it arises from his personal actions upon the external world.  Thus, to a QBist, Einstein's statement doesn't establish a thing.

The QBist translation of the EPR story invoking Tenets 2 and 3 looks like this.  Whereas the Bell scenario requires the tacit supposition of a third-person account of two distant measuring devices going click, click, click to even be posed  (see discussed in Section~\ref{TenetTwo}), the EPR scenario can be described from a first-person point-of-view from the outset.  Alice starts with an entangled quantum state for a bipartite system, one half of which might be far away.  This quantum state refers only to her beliefs about {\it her own potential experiences\/} should she take an action on either or both of these systems.  Suppose she takes an action on the system at A\@.  The consequence of her action on A causes her to update her quantum state for B\@.  So what?  What does it mean?  Does it mean something spooky happened at a distance?  No, it only tells of what {\it she\/} should expect should {\it she\/} take an action on B\@.  On the other hand, under the assumption of locality, how could she take an action on B if not to walk over to it?  To repeat, QBism is a strictly local account of quantum theory~\cite{Fuchs14}, so making an updated quantum-state assignment for B means nothing more than an updating of the agent's expectations for what might happen {\it to her\/} should she take an action upon system B.

And there goes poof to the alleged incompleteness, at least in the sense that EPR saw it.  Alice cannot take the complementary actions on A simultaneously, but even suspending all of quantum mechanics and imagining that she could, on the far-subjective Bayesian account of probability which QBism adopts, even ``probability equal to unity'' (as EPR put it) cannot tell nature what to do.  That does not sound like Bohr to my ears, but who knows?  Nonetheless, there is still a gem to be mined from Bohr's reply to EPR, and we turn to this next.

\subsection{Bohr's Phenomena, QBism's Experience}

\medskip

\begin{flushright}
\baselineskip=13pt
\parbox{4.0in}{\baselineskip=13pt\footnotesize
Notwithstanding all differences between the physical problems which have given rise to the development of relativity theory and quantum theory, respectively, a comparison of purely logical aspects of relativistic and complementary argumentation reveals striking similarities as regards {\bf the renunciation of the absolute significance of conventional physical attributes of objects}.}\\
\footnotesize --- Niels Bohr \cite[p.\ 64]{Bohr49-1}
\end{flushright}

In the Introduction to this paper, we stated that of all Bohr's extant thinking, what is perhaps most valuable for future inspiration to QBism is his notion of ``phenomenon.''  Let us let Bohr speak for himself with a few passages.

\bq\small
As a more appropriate way of expression I advocated the application
of the word phenomenon exclusively to refer to the observations
obtained under specified circumstances, including an account of the
whole experimental arrangement. In such terminology, the observational
problem is free of any special intricacy since, in actual experiments,
all observations are expressed by unambiguous statements referring,
for instance, to the registration of the point at which an electron
arrives at a photographic plate. Moreover, speaking in such a
way is just suited to emphasize that the appropriate physical interpretation
of the symbolic quantum-mechanical formalism amounts
only to predictions, of determinate or statistical character, pertaining
to individual phenomena appearing under conditions defined by classical
physical concepts. \cite[p.\ 64]{Bohr49-1}
\eq

\bq\small
\noindent
This crucial point \ldots\ implies the {\it impossibility of any
sharp separation between the behaviour of atomic objects and the
interaction with the measuring instruments which serve to define the
conditions under which the phenomena appear}. In fact, the individuality
of the typical quantum effects finds its proper expression
in the circumstance that any attempt of subdividing the phenomena
will demand a change in the experimental arrangement introducing
new possibilities of interaction between objects and measuring instruments
which in principle cannot be controlled.  \cite[pp.\ 39--40]{Bohr49-1}
\eq

\bq\small
\noindent This mutual exclusiveness of the experimental
conditions implies that the whole experimental arrangement must be
taken into account in a well-defined description of the phenomena.
The indivisibility of quantum phenomena finds its consequent\footnote{Be mindful of Bohr's use of the word ``consequent.''  Here is another instance of his usage, drawn from the correspondence with Pauli reported in Section \ref{PauliBohrDebate}.  Bohr writes, ``You are certainly right that Einstein is not consequent when speaking of the ideal of detached observer and neglecting his own wisdom of relativity \ldots,'' though that use appears after Pauli had already introduced the term in the previous letter.  My best guess is that Bohr meant to use ``consistent'' here, but was thrown off by the German ``konsequent'' (via Pauli) or the Danish ``konsekvent.''} expression
in the circumstance that every definable subdivision would
require a change of the experimental arrangement with the appearance
of new individual phenomena. \cite[p.\ 90]{Bohr55-1}
\eq

Of course, parts of these quotes cannot be right from the point-of-view of QBism.  Bohr says ``the observational problem is free of any special intricacy,'' but the question is, where does the measuring device stop and the observing agent begin?  For Bohr, the measuring devices can be (and should be!)\ excluded from a quantum description when they are heavy enough and capable of obtaining an irreversible mark upon them~\cite[pp.\ 91--92]{Bohr62-1}:
\bq\small
\noindent
The main argument was that unambiguous
communication of physical evidence demands that the experimental
arrangement as well as the recording of the observations be expressed
in common language, suitably refined by the vocabulary of classical
physics. In all actual experimentation this demand is fulfilled by using
as measuring instruments bodies like diaphragms, lenses and photographic
plates so large and heavy that, notwithstanding the decisive
role of the quantum of action for the stability and properties of such
bodies, all quantum effects can be disregarded in the account of their
position and motion.

While within the scope of classical physics we are dealing with an
idealization, according to which all phenomena can be arbitrarily subdivided,
and the interaction between the measuring instruments and
the object under observation neglected, or at any rate compensated for,
it was stressed that such interaction represents in quantum physics an
integral part of the phenomena, for which no separate account can be
given if the instruments shall serve the purpose of defining the conditions
under which the observations are obtained. In this connection
it must also be remembered that recording of observations ultimately
rests on the production of permanent marks on the measuring instruments,
such as the spot produced on a photographic plate by impact of a
photon or an electron. That such recording involves essentially irreversible
physical and chemical processes does not introduce any special
intricacy, but rather stresses the element of irreversibility implied in
the very concept of observation.
\eq

But there must be some point of contact between Bohr's phenomena and the quantum formalism, and presumably this is what the symbols $\hat\rho$ and $\{\hat D_j\}$ do (were Bohr to learn of positive operator-valued measures).  For Bohr, those symbols are names that ultimately signify ``classical physical explanation \ldots\ of what we have done and what we have learned.''  These symbols amount to objective facts that can be communicated between any two observers, though written stylistically in the symbolics of quantum theory.

On the other hand, for QBism these symbols are nothing more than proxies for a set of personal probability assignments.  Namely, the
\be
P(H_i) \quad\qquad \mbox{and} \quad\qquad P(D_j|H_i)
\ee
of Eqs.\ (\ref{Pass}) and (\ref{Punt}).  To put it starkly, of those quasi-realistic figures illustrating Bohr's 1949 ``Discussions with Einstein'' article---with their intentionally heavy bolts anchoring the solid devices to the table---QBism declares they are as diaphanous as any subjective probability assignment.  Their solidity is all illusion when it comes down to it.

So, where Bohr thinks that a separation between the agent and the measuring device is always securable in practice (and indeed necessary for the interpretation), QBism says, ``No, it is just the opposite.  If the agent were to go poof, the symbols $\hat\rho$ and $\{\hat D_j\}$ would go poof too.  Where's your classically-described preparation and measurement devices now?  Something is surely left behind, but it is not those things named by $\hat\rho$ and $\{\hat D_j\}$.''  Thus QBism chooses to dispense with the idea of the devices as things separate and foreign to the agent:  The agent and his devices are of one flesh.  And if it is so, then pace Bohr, there is indeed a very {\it special intricacy\/} to the observational problem in quantum theory.  It means everything Bohr says of ``atomic objects'' and ``measuring instruments''---when he speaks of the two together---becomes relevant instead to ``atomic objects'' and ``agent.''  For instance, parroting the above, ``This crucial point \ldots\ implies the {\it impossibility of any sharp separation between the behaviour of atomic objects and {\rm [the agent]} which serve to define the conditions under which the phenomena appear}.''  It is no wonder QBism invokes the concept of {\it experience\/} to express what a quantum measurement outcome {\it is}.

Indeed, there is something very compelling in Bohr's notion of phenomenon, suitably amended to be in terms of experience.  It is the part about, ``The indivisibility of quantum phenomena finds its consequent expression in the circumstance that every definable subdivision would require a change of the experimental arrangement with the appearance of new individual phenomena.''  Translating this one into the new setting---perhaps, ``The indivisibility of {\it experience\/} finds its consistent expression in the circumstance that every definable subdivision would require a change of the {\it agent's engagement with the world}, with the consequent\footnote{English version this time!} appearance of new individual {\it experience}.''---we get something that might well have come straight from the books of John Dewey \ldots\ writing style included!

Here is a sample from Dewey's ``The Postulate of Immediate Empiricism''~\cite{Dewey1905}:
\bq\small
For example, I start and am flustered by a noise heard. Empirically, that noise {\it is\/} fearsome; it {\it really\/} is, not merely phenomenally or subjectively so. That {\it is what\/} it is experienced as being. But, when I experience the noise as a {\it known\/} thing, I find it to be innocent of harm. It is the tapping of a shade against the window, owing to movements of the wind. The experience has changed; that is, the thing experienced has changed---not that an unreality has given place to a reality, nor that some transcendental (unexperienced) Reality has changed, not that truth has changed, but just and only the concrete reality experienced has changed. I now feel ashamed of my fright; and the noise as fearsome is changed to noise as a wind-curtain fact, and hence practically indifferent to my welfare. This is a change of experienced reality effected through the medium of cognition. The content of the latter experience cognitively regarded is doubtless {\it truer\/} than the content of the earlier; but it is in no sense more real. To call it truer, moreover, must, from the empirical standpoint, mean a concrete {\it difference\/} in actual things experienced. Again, in many cases, it is only in retrospect that the prior experience is cognitionally regarded at all. In such cases, it is only in regard to contrasted contents in the subsequent experience that the determination `truer' has force.

Perhaps some reader may now object that as matter of fact the entire experience {\it is\/} cognitive, but that the earlier parts of it are only imperfectly so, resulting in a phenomenon which is not real; while the latter part, being a more complete cognition, results in what is relatively, at least, more real. In short, a critic may say that, when I was frightened by the noise, I {\it knew\/} I was frightened; otherwise there would have been no experience at all. At this point, it is necessary to make a distinction so simple and yet so all-fundamental that I am afraid the reader will be inclined to pooh-pooh it away as a mere verbal distinction. But to see that {\it to the empiricist\/} this distinction is not verbal, but real, is the precondition of any understanding of him. The immediatist must, by his postulate, ask what is the fright experienced {\it as}. Is what is actually experienced, I-know-I-am-frightened, or I-{\it am}-frightened? I see absolutely no reason for claiming that the experience {\it must\/} be described by the former phrase. In all probability %(and all the empiricist logically needs is just one case of which this is true)
\ldots\ the experience is simply and just of fright-at-the-noise. Later one may (or may not) have an experience describable {\it as\/} I-know-I-am- (-or-was) and improperly or properly, frightened. But this is a different experience---that is, a different {\it thing}. %And if the critic goes on to urge that the person {\it `really'\/} must have known that he was frightened, I can only point out that the critic is shifting the venue.
\eq
Bohr's indivisibility of phenomena, in QBist hands, becomes Dewey's indivisibility of experience.  Or at least it becomes something more along those lines.  What is there to learn here?  Surely it is something big:  For now, it is not Dewey or James saying these things, but something compelled by the formal structure of quantum theory.

In fact, we have gotten here by all the toil and hard work in quantum foundations that came after Bohr's death.  What really shores up Bohr's pronouncement is the number of strong no-go theorems of so many varieties that---to the reasonable mind---indicate simply no hidden variables, no ontic quantum states, no nonlinear modifications to quantum theory, no picture of quantum theory that does not deeply involve the notion of experience or active, indecomposable events at its core.

Wheeler liked to say, ``Philosophy is too important to be left to the philosophers!''  I would not go that far, but there is something exciting in all of this.  One volunteers a philosophy, but one does not volunteer a physics.  A physics either flies in the world, or it falters and is eliminated by Darwinian selection.  That our most encompassing physical theory yet might lead to a philosophy once volunteered by temperament is a very powerful development.  The {\it forum\/} may perhaps finally take note as James suggested in his Lowell Lectures, for this philosophy now has some ``conventionally recognized reason'' on its side.

The case QBism makes before the forum is this.  What the quantum agent---the protagonist in the drama of any application of quantum theory---is ultimately doing is hitching a ride with a {\it new kind of ontology or metaphysic}.  An ontology of all-pervasive, {\it pan-creative}\footnote{This terminology is adapted from, but may not be identical to, Michel Weber's designation of Alfred North Whitehead's later ontology as a {\it pancreativism}~\cite{Weber2006,Weber2011}. } experience, not unakin to what William James was thinking of when he wrote his essay ``Does `Consciousness' Exist?''~\cite{James96a}:
\bq\small
My thesis is that if we start with the supposition that there is only one primal stuff or material in the world, a stuff of which everything is composed, and if we call that stuff `pure experience,' then knowing can easily be explained as a particular sort of relation towards one another into which portions of pure experience may enter.  The relation itself is a part of pure experience; one of its `terms' becomes the subject or bearer of the knowledge, the knower,$^*$ the other becomes the object known.  [$^*$In my {\sl Psychology\/} I have tried to show that we need no knower other than the `passing thought.']
\eq

The argument for a metaphysic along these lines comes from the very stringency of QBism's conception of quantum theory as a ``user's manual'' {\it any\/} agent can adopt to better cope with the world.  What happens when two users of quantum theory confront each other in the act of communication?  In QBism, the asking of a question to one's fellow is an action on the external world like any other quantum measurement:  One agent querying another means the one agent is taking an action on the other, now thought of as a physical system.  Just as with any action on any quantum system, the act catalyzes a new experience for the agent taking it---there is nothing new here.  Alice speaking to Bob catalyzes a new experience for her.  But we could take it all again from Bob's perspective:  Bob speaking to Alice catalyzes a new experience for him.  In communication the situation is symmetrical:  From the one perspective, Alice is an agent, but from the other she is a physical system; similarly for Bob.  When Alice and Bob are in communication, the category distinctions are symmetrical: Like with the Rubin vase, the best the eye can do is flit back and forth between the two formulations.

This suggests that the ontological abstracta of our decision-theoretic conception of quantum theory is neither the agent, nor the object, but something that can be polarized in one direction or the other depending upon the attending analysis one gives it.  William James's ``pure experience'' and Alfred North Whitehead's ``actual occasions'' have something of this character\footnote{These ontologies are in a spectrum of the so-called {\it neutral monisms}.  For a recent review, see Ref.~\cite{Atmanspacher14}.  However, as Ruth Anna Putnam has emphasized, James's own view (and incidentally QBism's, see Sec.\ VI of \cite{Fuchs10a}) might be better characterized as a `neutral pluralism' \cite{Putnam97}:
\bq
[A] key element of James's radical empiricism is his rejection of mind/matter dualism as well as its reduction to either materialism or idealism.  In its place, he offers---it is the title of one of his essays---a world of pure experience.  In that world consciousness {\it as an entity\/} does not exist.  But neither is consciousness a function of matter, for matter {\it as an entity\/} also does not exist.  Ultimately there are only pure experiences (and, perhaps, experienceables---that is a difficult interpretative question), experiences which only in retrospect are {\it taken\/} either as part of a stream of thought or as physical objects.  Although one is tempted to call this view a neutral monism, it is, in my opinion, more properly thought of as a neutral pluralism---neutral in not favoring either thought or matter, plural because [as James says] ``there is no {\it general\/} stuff of which experience at large is made.  There are as many stuffs as there are `natures' in the things experienced \ldots\ and save for time and space (and, if you like, for `being') there appears no universal element of which all things are made.''
\eq
}, but QBism has more:  It has the guidance of the quantum formalism for shedding light on the notion.  When a ``pure experience'' is polarized into an agent-object division quantum theory rears its head by telling us the normative relations the agent should strive to satisfy with his beliefs.  These normative constraints cannot be independent of the character of the underlying primal stuff.  The task before QBism then is to reverse engineer---to distill and make explicit why these normative constraints (not some others) are imposed on us in the first place.  This is a large part of what the technical research program outlined at the end of Section \ref{PauliBohrDebate} is about: The first steps toward a QBism 2.0!

\section{Toward the Future: Bohr's Continuing Inspiration}
\label{Future}

\medskip

\begin{flushright}
\baselineskip=13pt
\parbox{4.0in}{\baselineskip=13pt\footnotesize
It is difficult to escape asking a challenging question. Is the
entirety of existence, rather than being built on particles or fields
of force or multidimensional geometry, built upon billions upon
billions of elementary quantum phenomena, those elementary acts of
``observer-participancy,'' those most ethereal of all the entities
that have been forced upon us by the progress of science?
}\\
\footnotesize --- John Archibald Wheeler \cite{Wheeler82c} \\
\end{flushright}

Few people know it, but William James may have been the first person to utter the idea of a big bang.  He records this in his last book {\sl Some Problems of Philosophy}, written just before his death in August 1910 and published posthumously~\cite{James40}:
\bq\small
\noindent
It is a common belief that all particular beings have one origin and source, either in God, or in atoms all equally old.  There is no real novelty, it is believed, in the universe, the new things that appear having either been eternally prefigured in the absolute, or being results of the same {\it primordia rerum}, atoms, or monads, getting into new mixtures.  But the question of being is so obscure anyhow, that whether realities have burst into existence all at once, by a single `bang,' as it were; or whether they came piecemeal, and have different ages (so that real novelties may be leaking into our universe all the time), may here be left an open question \ldots
\eq
John Wheeler had wondered whether the two ideas might somehow be identified.  Is there any sense to it?  It is not the aim of this essay to go that far.  But like Wheeler, James was a man of sweeping ideas, and had he known enough physics, it would have been wonderful to talk with him about it.  It would have been wonderful to introduce him to QBism's evolution of Bohr's thought.

Beyond all else, James stood for the idea that the universe is ``on the make''---from beginning to end and everywhere in between.  What he sought was a vision of existence where {\it novelty\/} itself is the crucial ingredient.  The idea of a {\it fallible\/} world best fit his temperament.  He wanted a world that could grow and become complex, but as well potentially decline and even become desolate.  He thought such was what made life worth living~\cite{James1895}, especially if we all (along with everything else) had a part to play in such growth or decline.

Is the big bang here?  Perhaps the closest James ever came to giving an answer was this~\cite{James1907}:
\bq\small
Does our act then {\it create\/} the world's salvation so far as it
makes room for itself, so far as it leaps into the gap? Does it
create, not the whole world's salvation of course, but just so much
of this as itself covers of the world's extent?
\eq

\bq\small
Here I take the bull by the horns, and in spite of the whole crew of
rationalists and monists, of whatever brand they be, I ask {\it why
not?} Our acts, our turning-places, where we seem to ourselves to
make ourselves and grow, are the parts of the world to which we are
closest, the parts of which our knowledge is the most intimate and
complete. Why should we not take them at their facevalue? Why may
they not be the actual turning-places and growing-places which they
seem to be, of the world---why not the workshop of being, where we
catch fact in the making, so that nowhere may the world grow in any
other kind of way than this?

Irrational!\ we are told. How can new being come in local spots and
patches which add themselves or stay away at random, independently of
the rest? There must be a reason for our acts, and where in the last
resort can any reason be looked for save in the material pressure or
the logical compulsion of the total nature of the world? There can be
but one real agent of growth, or seeming growth, anywhere, and that
agent is the integral world itself. It may grow all-over, if growth
there be, but that single parts should grow {\it per se\/} is
irrational.

But if one talks of rationality---and of reasons for things, and
insists that they can't just come in spots, what {\it kind\/} of a
reason can there ultimately be why anything should come at all?
\eq
This idea of a world that comes in local spots and patches James called a {\it pluriverse}---a structure (if so it should be called) that would never allow itself a rationalist's unification into a single block.

If QBism has something to do for the future, it is to contribute the precision language of quantum theory to this grand vision of the world.  Who could say what's at stake any better than Will Durant already has?
\bq\small
\noindent The value of a [pluriverse], as compared with a universe, lies in this, that where there are cross-currents and warring forces our own strength and will may count and help decide the issue; it is a world where nothing is irrevocably settled, and all action matters. A monistic world is for us a dead world; in such a universe we carry out, willy-nilly, the parts assigned to us by an omnipotent deity or a primeval nebula; and not all our tears can wipe out one word of the eternal script. In a finished universe individuality is a delusion; ``in reality,'' the monist assures us, we are all bits of one mosaic substance. But in an unfinished world we can write some lines of the parts we play, and our choices mould in some measure the future in which we have to live. In such a world we can be free; it is a world of chance, and not of fate; everything is ``not quite''; and what we are or do may alter everything. \cite[p.\ 673]{Durant06}
\eq
If not as eloquently, at least maybe the physics of the future will say it with more precision.

It is good to know who your heroes in physics are, whether they be Bohr, Pauli, Wheeler, or even William James.  Their words should not be written in stone, but thought on from time to time to set one's agenda and give prod to the imagination.  Bohr remains a driving force in our quantum thinking and in the fresh field of quantum information.  QBism thanks him for that as much as anyone else.

\bigskip\bigskip
\noindent {\Large \bf Acknowledgements}\bigskip

%\section{Acknowledgments}

I am greatly indebted to Blake C. Stacey, the ``Mr.\ Atoz'' of Planet QBism.  Without his time portals into our history, this essay could not have been written---his knowledge of QBism is encyclopedic.  I also thank Ignacio Cirac for asking what is the difference between QBism and ``the'' Copenhagen interpretation, Berge Englert for saying there is none, and Lubo\v{s} Motl for calling attention to just how poor the scholarship on this subject can be in some corners~\cite{Motl1,Motl2}.  This research was supported in part by the Foundational Questions Institute Fund on the Physics of the Observer (grant FQXi-RFP-1612), a donor advised fund at the Silicon Valley Community Foundation.

\pagebreak

%\bigskip\bigskip
\noindent {\Large \bf Appendix: The Complete Notwithstanding}
\vspace{-12pt}

\bq\noindent\small
\begin{itemize}
\item[1925a:] ``Notwithstanding the formal nature of these suggestions, they exhibit a close connection with the spectral regularities disclosed by Land\'e's analysis.''  \cite[p.\ 44]{Bohr25-1}

\item[1925b:] ``Notwithstanding that the results thus obtained constitute an important step towards the above-mentioned programme of accounting for the properties of the elements solely on the basis of the atomic number, it must be remembered, however, that the results do not allow of a unique association with mechanical pictures.''  \cite[pp.\ 44--45]{Bohr25-1}

\item[1927a:] ``Notwithstanding the difficulties which, hence, are involved in the formulation of the quantum theory, it seems, as we shall see, that its essence may be expressed by the so-called quantum postulate, which attributes to any atomic process an essential discontinuity, or rather individuality, completely foreign to the classical theories and symbolized by Planck's constant of action.''  \cite[p.\ 53]{Bohr27-1}

\item[1927b:] ``Notwithstanding this contrast, however, a formal connection with the classical ideas could be obtained in the limit where the relative difference in the properties of neighbouring stationary states vanishes asymptotically and where in statistical applications the discontinuities may be disregarded. \cite[pp.\ 69--70]{Bohr27-1}

\item[1932a:] ``Notwithstanding the subtle character of the riddles of life, this problem has presented itself at every stage of science, the very essence of scientific explanation being the analysis of more complex phenomena into simpler ones.''  \cite[p.\ 3]{Bohr32-1}

\item[1932b:] ``Notwithstanding the greater complexity of the general problems of atomic mechanics, the lesson taught us by the analysis of the simpler light effects has been most important for this development.''  \cite[p.\ 6]{Bohr32-1}

\item[1932c:] ``Notwithstanding the fact that the multifarious biological phenomena are practically inexhaustible, an answer to this question can hardly be given without an examination of the meaning to be given to physical explanation still more penetrating than that to which the discovery of the quantum of action has already forced us.''  \cite[pp.\ 8--9]{Bohr32-1}

\item[1932d:] ``Notwithstanding the essential importance of the atomistic features, it is typical of biological research, however, that we can never control the external conditions to which any separate atom is subjected to the extent possible in the fundamental experiments of atomic physics.''  \cite[p.\ 10]{Bohr32-1}

\item[1932e:] ``Notwithstanding the intrinsic complexity of this problem, it is characteristic of our atom model that, owing to the predominance of the nuclear attraction over the mutual repulsion of the electrons in the inner region of the atom, we should expect a close resemblance between the R\"ontgen spectrum of an element and the spectrum emitted by the binding of a single electron to the nucleus.''  \cite[pp.\ 41--42]{Bohr32-2}

\item[1932f:] ``Notwithstanding the fundamental limitation of mechanical and electromagnetical ideas already emphasised, the essential reality of the results obtained in this way was also confirmed by the explanation of the remarkable selection rules, governing the appearance of spectral lines predicted by the combination principle, which was offered by correspondence arguments of the kind indicated in our discussion of the hydrogen spectrum.''  \cite[p.\ 44]{Bohr32-2}

\item[1932g:] ``Notwithstanding this formal similarity, the striking departure from classical ideas of statistics with which we have here to do presents, from the point of view of correspondence, an important difference in the cases of photons and of material particles like helium nuclei.''  \cite[p.\ 52]{Bohr32-2}

\item[1932h:] ``Notwithstanding the essentially new situation created by the discovery of the quantum of action, the characteristic feature with which we have here to do is not unfamiliar in atomic theory.''  \cite[p.\ 54]{Bohr32-2}

\item[1932i:] ``Notwithstanding its fertility, the attack on atomic problems in which the particle idea and the quantum of action are considered as independent foundations is of an essentially approximative character, since it does not allow of a rigorous fulfillment of the claim of relativistic invariance.''  \cite[p.\ 65]{Bohr32-3}

\item[1937a:] ``Notwithstanding the encouragement given to the pursuit of such inquiries by the great example of relativity theory which, just through the disclosure of unsuspected presuppositions for the unambiguous use of all physical concepts, opened new possibilities for the comprehension of apparently irreconcilable phenomena, we must realize that the situation met with in modern atomic theory is entirely unprecedented in the history of physical science.''  \cite[p.\ 19]{Bohr37-1}

\item[1937b:] ``Notwithstanding all differences, a certain analogy between the postulate of relativity and the point of view of complementarity can be seen in this, that according to the former the laws which in consequence of the finite velocity of light appear in different forms depending on the choice of the frame of reference, are equivalent to one another, whereas, according to the latter the results obtained by different measuring arrangements apparently contradictory because of the finite size of the quantum of action, are logically compatible.''  \cite[p.\ 85]{Bohr37-2}

\item[1938a:] ``Notwithstanding the great separation between our different branches of knowledge, the new lesson which has been impressed upon physicists regarding the caution with which all usual conventions must be applied as soon as we are not concerned with everyday experience may, indeed, be suited to remind us in a novel way of the dangers, well known to humanists, of judging from our own standpoint cultures developed within our societies.''  \cite[p.\ 23]{Bohr38-1}

\item[1938b:] ``Notwithstanding the admittedly practical necessity for most scientists to concentrate their efforts in special fields of research, science is, according to its aim of enlarging human understanding, essentially a unity.''  \cite[p.\ 92]{Bohr38-2}

\item[1938c:] ``Notwithstanding the adequacy of Einstein's idea of the photon to account for the exchange of energy and momentum in individual radiative processes, it was clear from the outset that the wealth of experience which had led to the acceptance of the wave-picture of light propagation put any revival of a simple corpuscular theory of radiation out of question.''  \cite[p.\ 95]{Bohr38-3}

\item[1938d:] ``Notwithstanding their great importance in illuminating typical aspects of atomic processes, the two kinds of quantum phenomena just discussed represent of course only limiting cases of special simplicity.''  \cite[p.\ 103]{Bohr38-3}

\item[1949a:] ``Notwithstanding its fertility, the idea of the photon implied a quite unforeseen dilemma, since any simple corpuscular picture of radiation would obviously be irreconcilable with interference effects, which present so essential an aspect of radiative phenomena, and which can be described only in terms of a wave picture.''  \cite[p.\ 34]{Bohr49-1}

\item[1949b:] ``Notwithstanding the renunciation of orbital pictures, Hamilton's canonical equations of mechanics are kept unaltered and Planck's constant enters only in the rules of commutation
$$
qp-pq=\sqrt{-1}\frac{h}{2\pi}
$$
holding for any set of conjugate variables $q$ and $p$.''  \cite[p.\ 38]{Bohr49-1}

\item[1949c:] ``Notwithstanding all novelty of approach, causal description is upheld in relativity theory within any given frame of reference, but in quantum theory the uncontrollable interaction between the objects and the measuring instruments forces us to a renunciation even in such a respect.''  \cite[p.\ 41]{Bohr49-1}

\item[1949d:] ``Notwithstanding the most suggestive confirmation of the soundness and wide scope of the quantum-mechanical way of description, Einstein nevertheless, in a following conversation with me, expressed a feeling of disquietude as regards the apparent lack of firmly laid down principles for the explanation of nature, in which all could agree.''  \cite[p.\ 56]{Bohr49-1}

\item[1949e:] ``Notwithstanding all differences between the physical problems which have given rise to the development of relativity theory and quantum theory, respectively, a comparison of purely logical aspects of relativistic and complementary argumentation reveals striking similarities as regards the renunciation of the absolute significance of conventional physical attributes of objects.''  \cite[p.\ 64]{Bohr49-1}

\item[1951a:] ``Notwithstanding the simplicity in such respects of our ideas concerning atomic constitution, it is equally clear that it is not possible, on the basis of the principles which have proved so fruitful in the description and comprehension of large scale physical phenomena, to account in detail for the properties of atoms.'' \cite[p.\ 151]{Bohr51-1}

\item[1953a:] ``Notwithstanding all differences in such respects, general epistemological features of objective description may be perceived in the attitude to ethical problems in the various cultures, and especially expressed in the religions.''  \cite[p.\ 159]{Bohr53-1}

\item[1954a:] ``Notwithstanding the astounding power of quantum mechanics, the radical departure from accustomed physical explanation, and especially the renunciation of the very idea of determinism, has given rise to doubts in the minds of many physicists and philosophers as to whether we are here dealing with a temporary expedient or are confronted with an irrevocable step as regards objective description.''  \cite[p.\ 72]{Bohr54-1}

\item[1954b:] ``Notwithstanding the inspiration required in all work of art, it may not be irreverent to remark that even at the climax of his work the artist relies on the common human foundation on which we stand.''  \cite[p.\ 79]{Bohr54-1}

\item[1956a:] ``Notwithstanding the measure to which it has been possible by familiar physical approach to further and utilize our knowledge about atoms, we have at the same time been confronted with unsuspected limitations of the ideas of classical physics demanding a revision of the foundation for the unambiguous application of some of our most elementary concepts.''  \cite[p.\ 171]{Bohr56-1}

\item[1958a:] ``Notwithstanding refinements of terminology due to accumulation of experimental evidence and developments of theoretical conceptions, all account of physical experience is, of course, ultimately based on common language, adapted to orientation in our surroundings and to tracing relationships between cause and effect.''  \cite[p.\ 1]{Bohr58-1}

\item[1958b:] ``Notwithstanding the power of quantum mechanics as a means of ordering an immense amount of evidence regarding atomic phenomena, its departure from accustomed demands of causal explanation has naturally given rise to the question whether we are here concerned with an exhaustive description of experience.''  \cite[p.\ 3]{Bohr58-1}

\item[1958c:] ``Notwithstanding all the difference in the typical situations to which the notions of relativity and complementarity apply, they present in epistemological respects far-reaching similarities.''  \cite[p.\ 6]{Bohr58-1}

\item[1960a:] ``Notwithstanding refinements of terminology due to accumulation of experimental evidence and developments of theoretical conceptions, all account of physical experience is, of course, ultimately based on common language, adapted to orientation in our surroundings and to tracing of relationships between cause and effect.''  \cite[p.\ 180]{Bohr60-1}

\item[1961a:] ``Notwithstanding the ever-increasing difficulties of postal communication, a steady correspondence with Rutherford was kept up.''  \cite[p.\ 50]{Bohr61-1}

\item[1961b:] ``Notwithstanding the remarkable analogy between essential features of atomic processes and classical resonance problems, it must indeed be taken into account that in wave mechanics we are dealing with functions which do not generally take real values, but demand the essential use of the symbol $\sqrt{-1}$ just as the matrices of quantum mechanics.''  \cite[pp.\ 56--57]{Bohr61-1}

\item[1961c:] ``Notwithstanding the help which mathematics has always offered for such a task, it must be realized that the very definition of mathematical symbols and operations rests on simple logical use of common language.''  \cite[p.\ 60]{Bohr61-1}

\item[1961d:] ``Notwithstanding the generalized significance of the superposition principle in quantum physics, an important guide for the closer study of observational problems was repeatedly found in Rayleigh's classic analysis of the inverse relation between the accuracy of the image-forming by microscopes and the resolving power of spectroscopic instruments.'' \cite[p.\ 61]{Bohr61-1}

\item[1962a:] ``Notwithstanding such general considerations, it appeared for a long time that the regulatory functions in living organisms, disclosed especially by studies of cell physiology and embryology, exhibited a finiteness so unfamiliar to ordinary physical and chemical experience as to point to the existence of fundamental biological laws without counterpart in the properties of inanimate matter studied under simple reproducible experimental conditions.''  \cite[p.\ 26]{Bohr62-1}

\item[1962b:] ``Notwithstanding the radical departure from deterministic pictorial description, with which we are here concerned, basic features of customary ideas of causality are upheld in the correspondence approach by referring the competing individual processes to a simple superposition of wave functions defined within a common space-time extension.'' \cite[p.\ 99]{Bohr62-2}

\end{itemize}
\eq

%%%%%%%%%%%%%%%%%%%%%%%%%%%%%%%%%%%% End Matter

\bibliographystyle{utphys}

\small

\bibliography{notwithstanding}

\end{document}